\documentclass[fleqn,useAMS,usenatbib]{mnras}
\usepackage{psfrag,color,epsfig}

\usepackage{graphicx}	% Including figure files
\usepackage{amsmath}	% Advanced maths commands
\usepackage{amssymb}	% Extra maths symbols
\usepackage{multicol}	% Multi-column entries in tables
\usepackage{bm}		% Bold maths symbols, including upright Greek
\usepackage{pdflscape}	% Landscape pages
\usepackage{multirow}
\usepackage{tabularx}  
\usepackage{caption}
\usepackage[T1]{fontenc}
\usepackage{ae,aecompl}

\def\xb{\bar{x}_{\rm \ion{H}{II}}}

\def\l{\mathcal L}

\title[Constraints on Excess Radio Background from LOFAR]{Tight Constraints on the Excess Radio Background at $z = 9.1$ from LOFAR} 
\author[R. Mondal, A. Fialkov et al.]{R. Mondal,$^{1,2}$\thanks{E-mail: rajesh@astro.su.se} A. Fialkov,$^{3}$\thanks{E-mail: afialkov@ast.cam.ac.uk} C. Fling,$^{1}$ I.T. Iliev,$^{1}$ R. Barkana,$^{4}$ B. Ciardi,$^5$   
\newauthor G. Mellema,$^2$ S. Zaroubi,$^{6,7,8}$ L.V.E Koopmans,$^8$ F. G. Mertens,$^8$ B. K. Gehlot,$^9$  
\newauthor R. Ghara,$^{2,6,7}$ A. Ghosh,$^{10}$ S. K. Giri,$^{2,11}$ A. Offringa$^{12}$ and V.N. Pandey$^{8,12}$\\
$^{1}$Astronomy Centre, Department of Physics and Astronomy, University of Sussex, Brighton BN1 9QH, UK\\
$^{2}$The Oskar Klein Centre, Department of Astronomy, Stockholm University, AlbaNova, SE-10691 Stockholm, Sweden\\
$^{3}$Institute of Astronomy, University of Cambridge, Madingley Road, Cambridge CB3 0HA, UK\\
$^4$Raymond and  Beverly Sackler School of Physics and Astronomy, Tel Aviv University, Tel Aviv 69978, Israel\\
$^5$Max-Planck Institute for Astrophysics, Karl-Schwarzschild-Str. 1, 85748 Garching, Germany\\
$^6$Department of Natural Sciences, The Open University of Israel, 1 University Road, PO Box 808, Ra'anana 4353701, Israel\\
$^7$Department of Physics, Technion, Haifa 32000, Israel\\
$^8$Kapteyn Astronomical Institute, University of Groningen, PO Box 800, 9700 AV Groningen, The Netherlands\\
$^9$School of Earth and Space Exploration, Arizona State University, 781 Terrace Mall, Tempe, AZ 85287, USA\\
$^{10}$Department of Physics, Banwarilal Bhalotia College, Asansol, West Bengal, India\\
$^{11}$Institute for Computational Science, University of Zurich, Winterthurerstrasse 190, 8057 Zurich, Switzerland\\
$^{12}$ASTRON -- the Netherlands Institute for Radio Astronomy, Oude Hoogeveensedijk 4, 7991 PD Dwingeloo, The Netherlands\\
}

\date{}
\pubyear{2018}

\begin{document}
\label{firstpage}
\pagerange{\pageref{firstpage}--\pageref{lastpage}}
\maketitle

%--------------------------------------------------------------------

\begin{abstract}
The ARCADE2 and LWA1 experiments  have claimed an excess over the Cosmic Microwave Background~(CMB) at low radio frequencies. If the cosmological high-redshift contribution to this radio background is between 0.1\% and 22\% of the CMB at 1.42\,GHz, it could explain the tentative EDGES Low-Band detection of the anomalously deep absorption in the 21-cm signal of neutral hydrogen. We use the upper limit on the 21-cm signal from the Epoch of Reionization ($z=9.1$)  based on 141\,hours of observations with LOFAR to evaluate the contribution of the high redshift Universe to the detected radio background. Marginalizing over astrophysical properties of star-forming halos, we find (at 95\% C.L.) that the cosmological radio background can be at most 9.6\% of the CMB at 1.42\,GHz. This limit rules out strong contribution of the high-redshift Universe to the ARCADE2 and LWA1 measurements. Even though LOFAR places limit on the extra radio background, excess of $0.1-9.6$\%  over the CMB (at 1.42\,GHz) is still allowed and could explain the EDGES Low-Band detection. We also constrain the thermal and ionization state of the gas at $z = 9.1$, and put limits on the properties of the first star-forming objects. We find that, in agreement with the limits from EDGES High-Band data, LOFAR data constrain scenarios with inefficient X-ray sources, and cases where the Universe was ionized by stars in massive halos only.
\end{abstract}

%In addition, we  constrain the thermal and ionization state of the gas at $z=9.1$. Finally, we put limits on the properties of the first star-forming objects in models with and without the excess radio background. } % showing that the mean temperature of the neutral gas is $31.2$\,K or higher (at 68\%). We put limits on the properties of the first star-forming objects, ruling out (all at 68\%) a star formation efficiency greater than 3\%, minimum mass of star-forming halos larger than $3.6\times 10^8\,$M$_{\sun}$ (equivalent to circular velocity larger than $26$\,km\,s$^{-1}$ at $z = 9.1$), X-ray luminosity per star formation rate lower than $6\times 10^{38}\,{\rm erg\,s}^{-1}M_{\sun}^{-1}\,{\rm yr}$, a CMB optical depth larger than $0.075$, and a mean free path of ionizing photons larger than 48 comoving Mpc. In addition, we present constraints on the thermal and ionization state of the intergalactic medium in models with and without the extra radio background. Finally, we show as well as on residual model parameters, astrophysical parameters with and without the standard scenario where the radio background is assumed to be the CMB. In this case the limits on the parameters of reionization are similar, while the constraints on the thermal history are weaker compared to the models with an extra radio background.

%--------------------------------------------------------------------

\section{Introduction}
\label{sec:intro}
Studies of the Epoch of Reionization~(EoR) and Cosmic Dawn are key to understanding early galaxy formation and the evolution of the Intergalactic Medium~(IGM) \citep[see e.g. reviews by][]{barkana01, furlanetto06, barkana18book, Mesingerbook}. Ionising properties of the high-redshift sources are currently largely constrained by the measurement of the electron scattering optical depth, $\tau$, estimated by Planck \citep[e.g.,][]{Planck:2018}, Ly-$\alpha$ damping wing absorption  in the spectra of the high-redshift quasars \citep[e.g.,][]{Greig:2017, Greig:2019} and Ly-$\alpha$ emission from Lyman Break galaxies \citep{Mason:2018}.  Accumulating evidence supports rapid and late reionization completed by $z\sim 6$ \citep[e.g.,][]{Weinberger:2019}; while galaxy surveys provide independent constraints on star formation out to $z\sim 10$ \citep[e.g., see][and references therein]{Behroozi:2019}. However, these observations do not constrain properties of the first population of star-forming objects such as their  star formation efficiency, feedback mechanisms that regulated primordial star formation, and the properties of the first  sources of heat (e.g., X-ray binaries). These properties can be probed using low-frequency radio observations of the redshifted 21-cm signal of neutral hydrogen \citep[e.g.,][]{pober14, Greig:2016, Singh:2017, Monsalve2018, Monsalve:2019}.

The 21-cm signal is produced by atomic hydrogen in the IGM. The hyper-fine splitting of the lowest energy level of a hydrogen atom gives rise to the rest-frame $\nu_{21} = 1.42$\,GHz radio signal with the equivalent wavelength of about 21\,cm \citep[see][for a recent review]{barkana18book, Mesingerbook}. Owing to its dependence on the underlying astrophysics and cosmology, this signal is a powerful tool to characterise the formation and the evolution of the first populations of astrophysical sources and, potentially, properties of dark matter, across cosmic time. Because the 21-cm signal is measured against the diffused radio background, usually assumed to be only the Cosmic Microwave Background~(CMB), this signal can also be used to constrain properties of any excess background radiation at low radio frequencies.

Recently, a detection of the global 21-cm signal from $z\sim 17$ was reported by the EDGES collaboration \citep{Bowman:2018}. The reported signal significantly deviates from  standard astrophysical models \citep[e.g.,][show a large set of viable 21-cm global signals varying astrophysical parameters in the broadest possible range]{Cohen:2017} and concerns about the signal being of cosmological origin have, therefore, been expressed \citep{Hills:2018, Sims:2019, Singh:2019, Bradley:2019, Spinelli:2019}. Despite these concerns, several theories  have been proposed to explain the  stronger than expected absorption, e.g., over-cooling of hydrogen gas by dark matter \citep{Barkana:2018}.  Alternatively, the existence of a new component of radio background at low radio frequencies in addition to the CMB could also lead to a deeper 21-cm absorption feature due to the stronger contrast between the temperatures of the background and the gas \citep[e.g.,][]{Bowman:2018, Feng:2018}. Astrophysical sources such as supernovae or accreting supermassive black holes \citep{Biermann:2014, EwallWice:2018, EwallWice:2019, jana2019, Mirocha:2019} could produce such an extra radio background. However, these sources would need to be several orders of magnitude more efficient in producing synchrotron radiation than corresponding sources at low redshifts (see \citealt{Sharma:2018} and \citealt{EwallWice:2019}), which is not very likely. An extra radio background can also be created by more exotic agents such as active neutrinos \citep{Chianese:2018}, dark matter  \citep{Fraser:2018, Pospelov:2018} or superconducting cosmic strings \citep{Brandenberger:2019}.  Interestingly, excess radio background at low radio frequencies was detected by the ARCADE2 collaboration at $3-90\,$GHz \citep{Fixsen:2011} as well as by LWA1  at $40-80\,$MHz \citep{Dowell:2018}. Specifically, the latter measurement shows that the excess can be fitted  by a power law with a spectral index of $ - 2.58\pm 0.05$ and a brightness temperature of $603^{+102}_{ - 92}\,$mK at the reference frequency $1.42\,$GHz. However, the nature of this excess is still debated  \citep{Subrahmanyan:2013}. 

Apart from the EDGES Low-Band, several other global signal experiments report upper limits. At Cosmic Dawn, an upper limit of $890$~mK on the amplitude of the 21-cm signal at $z\sim 20$ \citep{Bernardi:2016} was derived using the Large-Aperture Experiment to Detect the Dark Ages \citep[LEDA,][]{Price:2018}. At lower redshifts both the EDGES High-Band collaboration  \citep[$z \sim 6.5-14.8$,][]{Monsalve:2017} and the Shaped Antenna measurement of the background RAdio Spectrum \citep[SARAS2, $z\sim 6.1-11.9$,][]{Singh:2017} reported non-detection, which allowed to disfavour  astrophysical scenarios with negligible X-ray heating. Using the same astrophysical modelling as we employ here\footnote{Independent astrophysical constraints were obtained from the EDGES High-Band data \citep{Monsalve2018}  using a different set of models generated with {\sc 21cmFAST} \citep{Mesinger:2011}. }, the SARAS2 team ruled out 25 ''cold'' scenarios out of a set of 264 different signals compiled by \citet{Cohen:2017} at greater than 5$\sigma$ rejection significance \citep{Singh:2018}; while  \citet{Monsalve:2019} placed 68\% limits on the  X-ray heating efficiency of early sources   and other astrophysical parameters using EDGES High-Band data and 3.2 million models generated with the global signal emulator {\sc 21cmGEM} \citep{Cohen:2019}.

In parallel, interferometric radio arrays are placing upper limits on the fluctuations of the 21-cm signal, including the Low-Frequency Array \citep[LOFAR\footnote{\url{http://www.lofar.org}},][]{Patil2017, Gehlot:2019, LOFAR-EoR:2020}, the Murchison Widefield Array \citep[MWA\footnote{\url{http://www.haystack.mit.edu/ast/arrays/mwa}},][]{Beardsley:2016, Barry:2019, Li:2019, Trott:2020}, the Donald C.\ Backer Precision Array for Probing the Epoch of Reionization \citep[PAPER\footnote{\url{http://eor.berkeley.edu}},][]{Kolopanis:2019}, the Giant Metrewave Radio Telescope \citep[GMRT\footnote{\url{http://www.gmrt.ncra.tifr.res.in}},][]{Paciga:2013}, and the Owens Valley Radio Observatory Long Wavelength Array \citep[OVRO-LWA\footnote{\url{https://www.ovro.caltech.edu/}},][]{Eastwood:2019}.

The recently reported LOFAR measurements \citep{LOFAR-EoR:2020} are based on 141 hours of observations and are currently the tightest upper limits on the 21-cm power spectrum from $z=9.1$, making it possible to rule out scenarios of cold IGM. Using these data \citet{ghara2020} find a lower limit of 3.55 K (95\%) on the  gas temperatures   at $z=9.1$ in the case of their non-uniform scenario  with the CMB as the background radiation (for reference, gas temperature in an adiabatically expanding universe without astrophysical sources of heating is $2.1$ K at $z\sim 9.1$). In this paper we use the LOFAR upper limits to constrain any excess radio background. We also derive limits on astrophysical parameters and the properties of the IGM with and without the excess radio contribution. 
%Even though our standard-physics results are in  broad agreement with \citet{ghara2020}, who used somewhat different astrophysical modelling, we leave  a detailed comparison to future work. Finally, we  compare to the results of \citet{ghara2020}, which used somewhat different theoretical modelling assuming  standard astrophysical modelling in the absence of the excess radio background.

This  paper is structured as follows: In Section~\ref{sec:sim}, we describe the simulations used to generate the mock data sets of the 21-cm power spectra. In Section~\ref{sec:sim_setup} we describe the mock data set and the ranges of parameters probed. In Section~\ref{sec:method}, we discuss the statistical analysis employed to constrain the model parameters.   In  Section~\ref{sec:Res_radio}, we report our  constraints on the amplitude of the excess radio background and compare it to the values that could explain the EDGES Low Band detection. We also place limits on the thermal and ionization state of the gas at $z = 9.1$, and on the properties of the first star-forming objects. We provide a qualitative comparison with the results of \citet{ghara2020} in Section~\ref{Sec:compare}. Finally, we conclude in Section~\ref{sec:conc}.

\section{Simulated 21-cm signal}
\label{sec:sim}

\subsection{Theoretical Modelling}
\label{sec:thsim}

The 21-cm signal of neutral hydrogen observed against a background radiation of the brightness temperature $T_{\rm rad}$ (at 1.42 GHz at redshift $z$), depends on the processes of cosmic heating and ionization.   The brightness temperature of the 21-cm signal is given by:
\begin{equation}
T_{21}  = \frac{T_{\rm S}-T_{\rm rad}}{1+z}\left(1-e^{-\tau_{21}}\right),
\label{Eq:T}  
\end{equation}
where $T_{\rm S}$ is the spin temperature of the transition which at Cosmic-Dawn redshifts is coupled to the kinetic temperature of the gas, $T_{\rm gas}$, through Ly-$\alpha$ photons produced by stellar sources  \citep{Wouthuysen:1952, Field:1958}. The value of $\tau_{21}$  is the optical depth at redshift $z$ given by 
\begin{equation}
\tau_{21} = \frac{3h_{\rm pl}A_{10}c \lambda_{21}^2n_{\rm H}}{32\pi k_{B} T_{\rm S} (1+z) dv/dr}\,,
\end{equation}
where  $dv/dr = H(z)/(1+z)$ is the gradient of the line of sight component of the comoving velocity field, $H(z)$ is the Hubble rate at $z$, and $n_{\rm H}$ is the neutral hydrogen number density at $z$ which depends on the ionization fraction and is driven by both ultraviolet and X-ray photons.  The spin temperature encodes complex astrophysical dependencies and can be written as 
\begin{equation}
T_{\rm S} = \frac{1+x_C +x_{\alpha}}{T_{\rm rad}^{-1}+\left(x_C +x_{\alpha}\right)T_{\rm gas}^{-1}}\,,
\end{equation}
where $x_C$ is the collisional coupling coefficient and $x_{\alpha}$  is the Wouthuysen-Field coupling coefficient \citep{Wouthuysen:1952, Field:1958}. Both  $x_C$  and $x_{\alpha}$ depend on the value of $T_{\rm rad}$:
 \begin{equation}
 x_\alpha = \frac{4P_\alpha}{27 A_{10}}\frac{T_*}{T_{\rm rad}},
 \label{eq:xa}
 \end{equation}
 with $P_\alpha$ being the total rate (per atom) at which Ly-$\alpha$ photons are scattered within the gas and $T_*$ is the effective temperature of the 21-cm transition (0.068\,K). The collisional coupling coefficient is
\begin{equation}
x_C = \frac{n_i \kappa^i_{10}}{A_{10}}\frac{T_*}{T_{\rm rad}},
\label{eq:xc}
\end{equation} 
 where $\kappa^i_{10}$  is the rate coefficient for spin de-excitation in collisions with the
species of type $i$ of density $n_i$, where we sum over species $i$ \citep[see e.g., ][for a recent review]{Barkana:2016}.

\subsubsection{Radio Background}
Usually the CMB is assumed to be sole contributor to the background radiation and $T_{\rm rad} = T_{\rm CMB}(1+z)$, where $T_{\rm CMB}$ is the present day value of the CMB temperature, 2.725\,K.  However, as was mentioned in Section~\ref{sec:intro}, the anomalously strong  EDGES Low-Band signal has encouraged the development of alternative models in which the radio background is enhanced  \citep[e.g.,][]{Bowman:2018, Feng:2018}. Here we adopt a phenomenological global extra radio background with a synchrotron spectrum in agreement with observations by LWA1. The total radio background at redshift $z$ is then  given by
\begin{equation}
T_{\rm rad} = T_{\rm CMB}(1+z)\left[1+A_{\rm r}\left(\frac{\nu_{\rm obs}}{78~{\rm MHz}}\right)^{\beta}\right],
\label{Eq:Trad}
\end{equation}
where  $\nu_{\rm obs}$ is the observed frequency, $A_{\rm r}$ is the amplitude defined relative to the CMB temperature and calculated at the reference frequency of 78 MHz (which is the centre of the absorption trough reported by the EDGES collaboration) and $\beta = - 2.6$ is the spectral index (in agreement with the LWA1 observation). We vary the value of $A_{\rm r}$ between 0 and 400 at 78 MHz with the upper limit being close to the LWA1 limit and corresponds to 21\% of the CMB at 1.42 GHz. All values of $A_{\rm r}$ between  $1.9$ (equivalent to 0.1\% of the CMB at 1.42 GHz) and 400 were shown to explain the EDGES Low detection \citep[for a tuned set of astrophysical parameters; see more details of the modelling in][]{Fialkov:2019}.

\subsubsection{Astrophysical Parameters}
Astrophysical processes affect the 21-cm signal by regulating the thermal and ionization states of the gas. In our modelling, we account for the effect of radiation (Ly-$\alpha$, Lyman-Werner, X-ray and ionizing radiation) produced by stars and stellar remnants on the 21-cm signal \citep{Visbal:2012, Fialkov:2013, fialkov14, Fialkov2014, Cohen:2016}.  The process of star formation is parameterized by two parameters. The first one is the value of circular velocity of dark matter halos, $V_c$, which is varied between 4.2\,km\,s$^{-1}$ (molecular hydrogen cooling limit, corresponding to the dark matter mass of M$_{\rm h}=1.5\times 10^{6}\,$M$_{\sun}$ at $z=9.1$) and 100\,km\,s$^{-1}$ (M$_{\rm h}=2\times 10^{10}\,$M$_{\sun}$ at $z=9.1$). The high values of $V_c$ implicitly take into account various chemical and mechanical feedback effects (e.g., the supernovae feedback which is expected to expel gas from small halos thus rising the threshold mass for star formation) which we do not include explicitly. Cooling of gas via molecular hydrogen cooling channel, and subsequent star formation, happens in small halos of circular velocity 4.2\,km\,s$^{-1} < V_c < 16.5$ \,km\,s$^{-1}$ (M$_{\rm h}\sim 10^5 -10^7\,$M$_{\sun}$). Abundance of molecular hydrogen is suppressed by Lyman-Werner~(LW) radiation \citep{Haiman:1997, Fialkov:2013}. Additional inhomogeneous suppression is introduced by the relative velocity between dark matter and baryons, $v_{\rm bc}$ \citep{Tseliakhovich:2010}, which imprints the pattern of Baryon Acoustic Oscillations~(BAO) in the 21-cm signal \citep{Dalal:2010, Maio:2011, Visbal:2012}. Higher mass halos ($V_c>16.5$\,km\,s$^{-1}$) form stars owing to atomic hydrogen cooling and are sensitive to neither the LW feedback nor to the effect of $v_{\rm bc}$, but are affected by photoheating feedback \citep{Sobacchi:2013, Cohen:2016,2018MNRAS.473...38S}.  The second parameter  is the star formation efficiency, $f_*$, defined as the amount of gas in halos that is converted into stars, which we vary in the range $f_* = 0.1$\% to 50\%.  Star formation in molecular cooling halos is assumed to be less efficient, which is implemented via a gradual low-mass cutoff \citep[see][for more details]{Cohen:2017}. The broad considered range in the values of $f_*$ is due to the lack of direct observations at high redshifts. Existing simulations of primordial star formation, although in general predict low values of $f_*$, show a large scatter in this parameter  \citep[e.g.,][]{Wise:2014, Oshea:2015, Xu:2016}. 

X-ray sources re-heat and mildly re-ionise the gas after the period of adiabatic cooling. Population synthesis simulations \citep{fragos13} calibrated to low-redshift observations of X-ray binaries \citep[e.g.,][]{Mineo:2012} suggest that high-mass X-ray binaries dominate the total X-ray budget at redshifts above $z\sim 6$. Here we rely on this result and assume hard X-ray spectral energy distribution~(SED) typical for a population of high-mass X-ray binaries at high redshifts  \citep[a complex function of X-ray energy with a peak at $\sim 2$\,keV adopted from][]{fragos13}. Another free parameter related to X-ray sources is the total X-ray luminosity, $L_X$. Observations of X-ray binaries in the local universe find a strong correlation between  $L_X$  and the star formation rate \citep[e.g.,][]{Lehmer:2010, Mineo:2012}. We adopt this dependence 
\begin{equation}
L_X/{\rm SFR} = 3\times 10^{40}f_X\,{\rm erg\,s}^{-1}M_{\sun}^{-1}\,{\rm yr},
\label{Eq:fX}
\end{equation}
introducing a normalisation constant, $f_X$, which accounts for a possible change in X-ray efficiency at high redshifts. Here we explore the wide range $f_X = 10^{-6}-100$. A value $f_X =1 $ yields $L_X$ normalized to observations of X-ray binaries found in low metallicity regions today \citep[see][and references therein]{fragos13}. Values of $f_X \gtrsim 100$ are unlikely \citep{Fialkov:2017} as such a population would saturate the unresolved X-ray background observed by the {\it Chandra}  X-ray Observatory \citep{Cappelluti:2012, Lehmer:2012}, while values $f_X \lesssim 10^{-6}$ contribute negligible X-ray heating. We note that in this paper we ignore the ($\sim 10\%$) effect of the CMB heating \citep{venumadhav18} and the effect of Ly-$\alpha$ heating \citep[e.g.,][]{chuzhoy07, ghara19}.

In our simulations the effects of ionizing radiation (ultraviolet radiation from stars) are defined by two parameters: the mean free path of ionizing photons, $R_{\rm mfp} = 10-70$ comoving Mpc, and the ionizing efficiency of sources, $\zeta$, which is tuned  to yield the CMB optical depth $\tau$ in the range between 0.045 and 0.1. For a fixed set of astrophysical parameters, either $\zeta$ or $\tau$ can be used \citep[for more details on the relation between  $\zeta$ and $\tau$ see][]{Cohen:2019}. Here we choose to use the latter as it is directly probed by the CMB experiments.  The latest values of $\tau$ measured by the {\it Planck} satellite are $\tau = 0.054\pm 0.007$  \citep[e.g.,][]{Planck:2018}. However, because in this paper we focus on the  constraints at $z\sim 9$, we explore a broader range of values ($0.045-0.1$) including higher values of $\tau$ which can be constrained by the LOFAR data.

\subsection{Simulation Setup}
We use a hybrid computational framework\footnote{Our code has similar architecture to  the publicly available 21cmFAST code of \citet{Mesinger:2011}, but the implementation is completely independent.}  to estimate the evolution of  the large scale 21-cm signal \citep{Visbal:2012, fialkov14, Fialkov2014, Cohen:2017, Fialkov:2019}. The code takes into account all the physics specified in above. Processes on scales below the resolution scale of 3 comoving Mpc\footnote{The resolution of our simulation, 3 comoving Mpc, is motivated by the coherence scale of the relative velocity between dark matter and gas, $v_{\rm bc}$ \citep{Tseliakhovich:2010}} (such as star formation, LW and photoheating feedback effects, effects of $v_{\rm bc}$) are implemented using sub-grid prescriptions. Radiation produced by stars and stellar remnants is propagated accounting for the effects of redshift on the energy of the photons and absorption in the IGM. Reionization is implemented using an excursion set formalism \citep{furlanetto04a}. Astrophysical parameters ($f_*,~V_{\rm c},~f_{\rm X},~\tau,~R_{\rm mfp},~A_{\rm r}$, and the spectral energy distribution of X-ray photons) are received as an input. The code generates cubes of the 21-cm signal at every redshift along with the temperature of the neutral IGM, ionization state, intensity of the Ly-$\alpha$ and LW backgrounds. The comoving volume of each simulation box is 384$^3\,$Mpc$^3$. The simulation is run from $z = 60$ to $z=6$.

We do not vary cosmological parameters with the exception of $\tau$. The values of other cosmological parameters are fixed to the values reported by the {\it Planck} collaboration \citep{Planck:2014}.

\section{Mock data sets and parameter sets}
\label{sec:sim_setup}

% \begin{enumerate}
%\item {\it Excess-background} models in which the radio background at low frequencies is boosted with respect to the CMB (Eq.~\ref{Eq:Trad}). We have accumulated 7702 such models with the parameters within the ranges specified above.
%\item {\it Standard} models  where the background radiation is assumed to be the CMB. We have accumulated data from 14247 full simulations. The standard case is a limit of the excess-background case with $A_{\rm r} =0$.  
%\end{enumerate}

Using the framework described in the previous section, we run a total of 23972 simulations varying the  astrophysical and background parameters in the ranges outlined above. A total of 7702 of these models have a boosted radio background with respect to the CMB and are referred to as the  {\it excess-background} models,  while the remainder are reference standard models with $A_{\rm r} = 0$ (used as a separate data set). All these models were generated using the same set of initial conditions for the distribution and velocities of dark matter and baryons  (the {\it fiducial} IC).

For each simulation, we calculate the values of the spherically averaged binned 21-cm power spectra $P(k_c)$, where $k_c$ is the centre of a wave-number bin chosen by \citet{LOFAR-EoR:2020}. The power spectrum is averaged over redshifts $z=8.7-9.6$ (to account for the LOFAR bandwidth), and binned over wave-numbers in agreement with the LOFAR observational setup (see Table~\ref{tab:LOFAR-data} for the details of the wave-number binning). From each simulation, we also extract: the mean temperature of the gas in neutral regions at $z = 9.1$, $T_{\rm gas}$, the mean ionization fraction  at $z = 9.1$, $\xb$, the redshift at which the ionization fraction (of volume) is $50\%$, $z_{\rm re}$, and the duration of reionization $\Delta z$, defined as the redshift range between the epoch when the mean ionization fraction was 90\% and 10\%.  

\begin{table}
\begin{center}
\caption{Summary of LOFAR measurements directly taken from Table~4 of \citet{LOFAR-EoR:2020}. From left to right: central mode of each bin in units of $h\,{\rm Mpc^{-1}}$, the extent of each $k$ bin, spherically averaged power spectrum in each bin, $1 - \sigma$ error in the binned power spectrum.}
\label{tab:LOFAR-data}
\begin{tabular}{cccc}
\hline
$k_{\rm c}$ & $k_1-k_2$ & $\Delta_{21}^2$ & $\Delta_{\rm 21,\,err}^2$\\
$(h\,{\rm Mpc^{-1}})$ & $(h\,{\rm Mpc^{-1}})$ & $({\rm mK}^2)$ & $({\rm mK}^2)$\\
\hline
0.075 & $0.061-0.082$  & 3476 & 916\\
0.100 & $0.082-0.111$ & 9065 & 1155\\
0.133 &  $0.111-0.150$ & 20211 & 1598\\
0.179  &  $0.150-0.203$ & 55603 &  2684 \\
0.238  & $0.203- 0.274$ & 128842  & 4097\\
0.319  & $0.274-0.370$ &  255292 &  7727 \\
0.432 & $0.370-0.500$ & 441200 & 12778 \\
\hline
\end{tabular}
\end{center}
\end{table}

%The sampling of the parameter space was done irregularly. In the standard case the part of the astrophysical parameter space that can be constrained by LOFAR is small. In order to make sure that we are not missing important cases, we have first sampled this region over a dense grid; we then probed this part of the parameter space randomly to allow for a more efficient training of neural networks (as discussed below). Other regions of the parameter space were probed randomly and less dense. In the excess-background case, the part of the parameter space constrained by LOFAR is much larger than in the standard case. Therefore,  a smaller amount of models was needed  for efficient training of neural networks. The parameter space was probed randomly.

\subsection{Sample Variance} 
\label{sec:bias}

\begin{figure}
\includegraphics[width=3.3in]{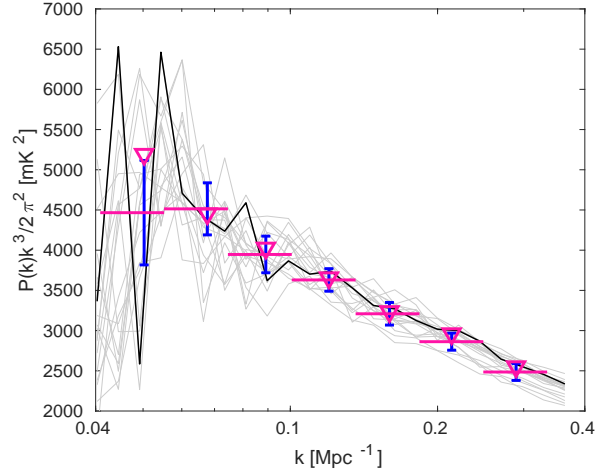}
\caption{Simulated power spectrum as a function of the comoving wavenumber binned in redshift  over the range $z=8.7-9.6$.  Solid black line is the result of a simulation with our fiducial set  of initial conditions, grey lines are the results of 17 additional runs with other sets of initial conditions but same astrophysical and cosmological parameters ($f_* = 45\%$, $V_c = 100\,$km s$^{-1}$, $R_{\rm mfp} = 55\,$Mpc, $\tau = 0.0738$, $f_X = 10^{-4}$, $A_{\rm r}=0$). Triangles mark the binned power spectra for our fiducial IC in the seven LOFAR bins; magenta horizontal lines show the extent of each wave-number  bin; blue error bars are the $1-\sigma$ variation in the binned power spectrum calculated from 18 realizations of the initial conditions. For the selected astrophysical scenario the deviations of the binned power spectra calculated from the fiducial set of IC from the ensemble mean are: 1.1$\sigma$, 0.19$\sigma$, 0.35$\sigma$, 0.17$\sigma$, 0.19$\sigma$, 0.9$\sigma$, and 0.6$\sigma$ (listed from the lowest to the highest wave-number). The corresponding values of $b_{\rm SV}$ (see the text) are 0.86, 1.01, 0.98, 0.99, 0.99, 0.97, 0.97 respectively.
%Model B (Right): $f_* = 30\%$, $V_c = 72.5$\,km/s, $R_{\rm mfp} = 50$ Mpc, $\zeta = 470$, $f_X = 0.01$. For this model the deviation from the mean of the lowest $k$-bin PS is 1.69$\sigma$ (and 0.1, 0.2 $\sigma$ in 2 and 3 bins respectively).
}
\label{Fig:CV}
\end{figure}

The lowest wave-number observed by \cite{LOFAR-EoR:2020} with LOFAR is $k_c = 0.075\,h\,$Mpc$^{-1}$, which corresponds to the scale of $\sim 125$ comoving Mpc and is a significant fraction of the size of our simulation box (384\,Mpc). Therefore, power spectrum in the lowest $k$-bin  is subject to statistical fluctuations due to {\it sample variance}, as is shown in  Figure~\ref{Fig:CV}.  For the set of initial conditions that we used to generate the entire data-set (our {\it fiducial} IC) the bin-averaged power spectrum in the lowest $k$-bin is 1.1$\sigma$ away from the mean calculated over 18 realizations. We correct for this systematic offset by introducing a bias factor.

We perform an auxiliary suite of simulations to systematically estimate the effect of sample variance. For each set of astrophysical parameters out of 360  selected  combinations\footnote{The astrophysical parameters were selected such that the power spectra at $k_c =0.075\,h\,$Mpc$^{-1}$ are close to the LOFAR measurements by \citet{LOFAR-EoR:2020}. This was done to ensure high precision in the testable range.},  10 simulations with different initial conditions,  including the fiducial set, were performed. The bias in the binned power spectrum was subsequently calculated for every $k$-bin as the ratio of the binned power spectrum averaged over 10 realizations to the one derived from the fiducial set:
\begin{equation}
b_{\rm SV}(k_c) = \frac{\bar P(k_c)}{P_{\rm fiducial}(k_c)}.
\end{equation}

We find that at $z = 9.1$ (close to the mid-point of reionization for the models that can be constrained by LOFAR in the standard case) the bias varies as a function of the reionization parameters $\tau$ and $R_{\rm mfp}$, while it has a very weak dependence on the rest of the parameters ($V_c$, $f_*$, $f_X$ and $A_{\rm r}$). We jointly fit the bias as a second order polynomial in $\tau$ times a linear function of $R_{\rm mfp}$. Because the entire data set described in Section~\ref{sec:sim_setup} was created using the fiducial IC set, we apply the corresponding parameter-dependent and $k_c$-dependent bias factor to all the simulated results to compensate for the effect of sample variance. Multiplying by the bias factor is essentially equivalent to averaging over 10 simulations.

Furthermore,  we fit the variation in the simulated power in each bin ($\sigma_{\rm SV,sim}(k_c)$, blue error bars in Figure~\ref{Fig:CV}), as a function of astrophysical parameters.  We find that  the fractional standard deviation, $\sigma_{\rm SV,sim}(k_c)/P_{\rm fiducial}(k_c)$, can be fitted with a quadratic function of  $\tau$ times a linear function of $R_{\rm mfp}$,  similarly to $b_{\rm SV}(k_c)$. The variation due to sample variance has a very weak dependence on  $V_c$, $f_*$, $f_X$ and $A_{\rm r}$. The error in the power spectrum (after it has been corrected by the bias factor) is then given by $\sigma_{\rm SV,sim}(k_c)/\sqrt{10}$.

%We also check whether the size of our simulation, $V_{\rm sim}$, matches the observed volume $V_{\rm LOFAR}$. A mismatch in the volumes of the simulated and the observed fields would lead to an erroneous estimate of the statistical error. The sample variance should scale as a square root of the number of modes, while the number of modes is proportional to the volume giving $\sigma_{\rm SV,sim}(k_c)\propto 1/\sqrt{V_{\rm sim}}$. {\color{magenta} Because the simulation volume is slightly bigger than the one observed by LOFAR, $V_{\rm LOFAR}$, the value of sample variance calculated at a given wavenumber $k$  using our simulation will be slightly smaller than if we simulated the actual LOFAR volume and used it to calculate sample variance at $k$.} Therefore, we need to scale sample variance with respect to the size of the LOFAR field,  multiplying  $\sigma_{\rm SV,sim}(k_c)$ by the correction factor $\sqrt{V_{\rm sim}/V_{\rm LOFAR}}$. The LOFAR field has a total volume of $\sim 92\times 10^6$ comoving Mpc$^3$ (it is 447 comoving Mpc $h^{-1}$ on a side, corresponding to 4$^\circ$, and has a depth of 139 comoving Mpc $h^{-1}$, corresponding to 11.5 MHz). Therefore, for our simulation box of 384$^3$ Mpc$^3$ we find that the volume correction factor is 0.78. 

Finally, in order to account for theoretical uncertainty in modelling\footnote{The values of the 21-cm signal generated by the numerical simulation are subject to uncertainty. This is because some of the effects of order $\sim(1+\delta$), where $\delta$ is the stochastic dimensionless perturbation of the density field, have not been taken into account. For example, at the moment we assume linear growth of structure on large scales ($>3$ Mpc).}, we impose  a lower limit of  10\% on the relative error of the power spectrum of each individual simulation \citep{ghara2020}. In the total error budget of the corrected power spectrum this error should also be divided by $\sqrt{10}$.

The total theoretical parameter-dependent error in the binned power spectrum is, thus, given by
\begin{equation}
\sigma_{\rm th}(k_c) = \sqrt{\frac{\left[0.1\times \Delta_{\rm th}^2(k_c)\right]^2 +  \sigma^2_{\rm SV,sim}(k_c)}{10}}\,,
\end{equation}
where $\Delta_{\rm th}^2(k_c) = P_{\rm fiducial}(k_c)k_c^3/\left(2\pi^2\right)$ is the calculated power spectrum in mK$^2$ units.
 
\subsection{Binning over the Model Parameters}
Our goal is to derive constraints on the excess radio background and also explore implications for the rest of the  model parameters, as well as for the thermal and ionization states of the IGM. Based on the value of the power spectrum for each set of model parameters, we  evaluate the likelihood of each point in the parameter space $\vec\theta$ as described in the next Section. We, therefore, need to bin the parameter space $\vec\theta$ and calculate the binned power spectra $\Delta_{\rm th}^2(k_c,\vec\theta)$ and the corresponding theoretical error $\sigma_{\rm th}(k_c,\vec\theta)$. To remind the reader, $\Delta_{\rm th}^2(k_c,\vec\theta)$ and $\sigma_{\rm th}(k_c,\vec\theta)$ are binned in redshift, wave-number and $\vec\theta$.

We explore two distinct  sets of the parameter spaces with $\vec\theta$ defined as either the model parameters  $\vec\theta = [f_*,\,V_c,\,f_X,\,\tau,\,R_{\rm mfp},\,A_{\rm r}]$ or the derived IGM quantities $\vec\theta= [T_{\rm gas},\,\xb,\,z_{\rm re},\,\Delta z]$ that describe the state of the IGM. The range of each parameter is divided into 10 equally spaced bins, and each bin is tagged by the bin-averaged value of relevant  parameters. Due to the large ranges, the binning is logarithmic for $f_*$, $V_c$,  $f_X$, $A_{\rm r}$ and $T_{\rm gas}$, and linear for $\tau$,  $R_{\rm mfp}$, $\xb$, $z_{\rm re}$ and  $\Delta z$. We assume flat priors on each of the parameters across the entire allowed range (see Section~\ref{sec:sim}):  $0.001 \le f_* \le 0.5$, 4.2 \,km\,s$^{-1} \le V_c \le 100$\,km\,s$^{-1}$, $10^{-6} \le f_X \le 100$, $0.045 \le \tau \le 0.1$, $10 \le R_{\rm mfp} \le 70$ comoving Mpc and zero outside these ranges. In the standard case $A_{\rm r} =0$ and in the excess background case, we vary $0.2 \le A_{\rm r} \le 400$ (thus covering the range 0.01\%-21\% of the CMB at 1.42 GHz). The priors on $[T_{\rm gas},\,\xb,\,z_{\rm re},\,\Delta z]$ are defined based on the  ranges of these  parameters found in our simulations: 2.2 \,K $\le T_{\rm gas} \le 400$\,K (the lower limit is close to the temperature of the gas in an adiabatically expanding universe which is $\sim 2.1$ K at $z=9.1$), $0.02 \le \xb \le 1.00$, $6 \le z_{\rm re} \le 10$ and $2 \le \Delta z \le 5$, and zero outside these ranges.

For $\vec\theta = [f_*,\,V_c,\,f_X,\,\tau,\,R_{\rm mfp},\,A_{\rm r}]$ this binning gives rise to $10^5$ bins in the standard case and $10^6$ bins in the excess-background case; for the IGM parameters $\vec\theta= [T_{\rm gas},\,\xb,\,z_{\rm re},\,\Delta z]$ there are $10^4$ bins in each case. Due to the relatively small number of models, not all bins are populated. To solve this issue, we use the model sets to train Artificial Neural Networks~(ANN, see Appendix~\ref{ann} for details) and use that to construct an emulator \citep[similar approach has been taken by][]{aviad2019, Monsalve:2019, Nicholas2017}, which we then use to interpolate the empty bins.

\section{Statistical Analysis Methodology}
\label{sec:method}
In general, the 21-cm signal is expected to be a non-Gaussian field \citep{bharadwaj05a,2006MNRAS.372..679M,mondal15} and the non-Gaussian effects will play a significant role in the error estimates of 21-cm power spectrum \citep{mondal16,mondal17}. In addition, the data in LOFAR bins are slightly correlated due to the finite station size. Therefore, the power spectrum error-covariance matrix is expected to be non-diagonal. However, in reality bins show very weak correlation because the bins are chosen relatively wide compared to the footprint of a LOFAR station which acts as a spatial convolution kernel. With minimal error, we can therefore assume that the bins are uncorrelated and the covariance matrix is diagonal. The probability of a model (tagged by $\vec\theta$)  given data can then be written as a  product of the probabilities in each individual wave-number bin $k_c \in k_i$. In addition, because of the bin-averaging and large scales considered, we can  assume that the signal is close to a Gaussian random field.

%\begin{equation}
%\l[\vec\theta |\Delta_{21}^2(k_i)] = \frac{1}{\sqrt{2\pi}\sigma(k_i) }\exp \left[-\frac{\{\Delta_{21}^2(k_i)- P_{\rm th}(k_i,\vec\theta)\}^2}{2\sigma^2(k_i)} \right],
%\label{eq:like}
%\end{equation}
%which is analogous to equation 11.20 of \citealt{dodelson_b03}. Here, $\Delta_{21}^2(k)$ is the measured power spectrum with uncertainty $\Delta_{\rm 21,\,err}^2$ (listed in Table~\ref{tab:LOFAR-data}). The total variance in the bin is given b

The LOFAR measurements reported by \citet{LOFAR-EoR:2020} are upper limits. Therefore, following \citet{ghara2020}, we can represent the probability of a model given the observed power spectrum values using the error function: 
\begin{equation}
\l(\vec\theta) = \prod_{i} \frac{1}{2} \left[ 1 + {\rm erf}\left\{\frac{\Delta_{21}^2(k_i) - \Delta_{\rm th}^2(k_i,\vec\theta)}{\sqrt{2} \sigma(k_i,\vec\theta)}\right\}\right]\,,
\label{eq:like1}
\end{equation}
where $\Delta_{21}^2(k_i)$ is the measured power spectrum in the $i$-th $k_c$ bin with uncertainty $\Delta_{\rm 21,\,err}^2(k_i)$ listed in Table~\ref{tab:LOFAR-data}. The total variance in the bin is given by
\begin{equation}
\sigma(k_i,\vec\theta) = \sqrt{[\sigma_{\rm th}(k_i,\vec\theta)]^2 + [\Delta^2_{\rm 21,\,err}(k_i)]^2}.
\end{equation}
According to this definition, the probability of a model is close to unity when its power spectrum at $z= 9.1$ is less than $[\Delta_{21}^2(k_i)- \sigma(k_i,\vec\theta)]$ for all $k_i$,  and the probability is close to zero when  $\Delta_{\rm th}^2(k_i,\vec\theta)$ is greater than $ [\Delta_{21}^2(k_i) + \sigma(k_i,\vec\theta)]$ for any $k_i$. 

\begin{figure*}
\includegraphics[width=3.2in]{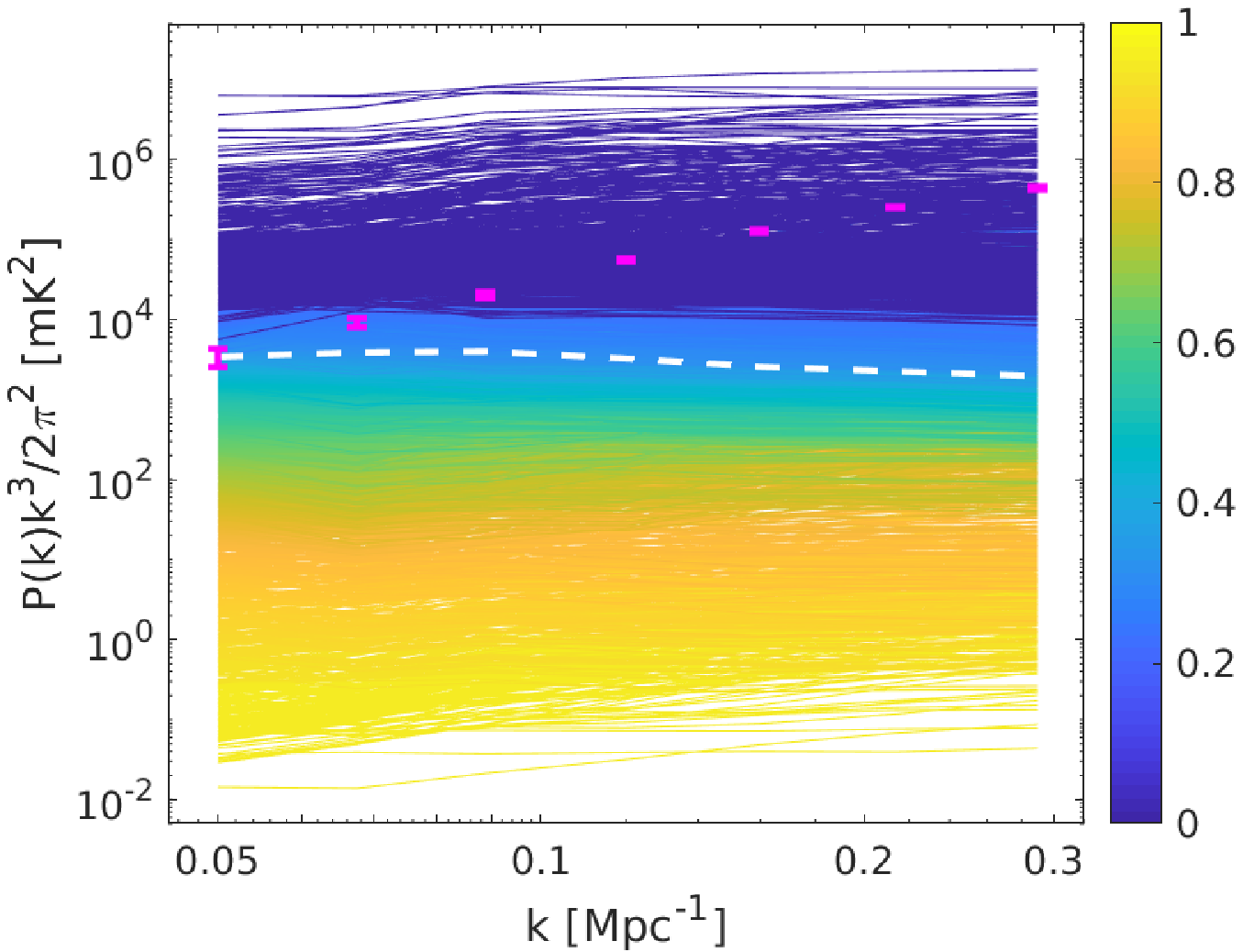}
\includegraphics[width=3.2in]{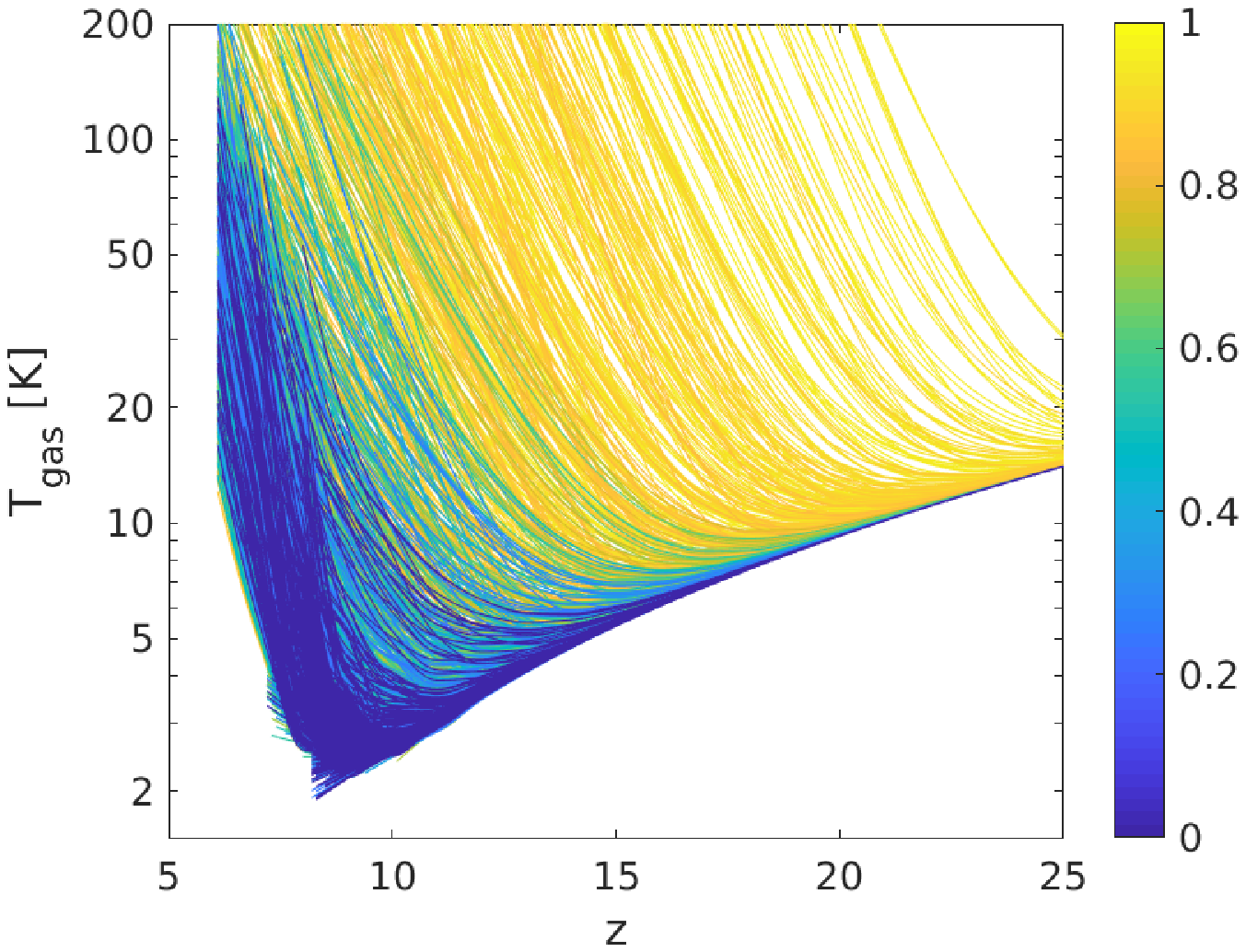}
\caption{We show the excess-background models colour-coded with respect to the probability that the data is consistent with the model (Eq.~\ref{eq:like1}) as is indicated  on the colour bar. Left: Binned power spectra vs wavenumber (in units of Mpc$^{-1}$, where we have assumed $h = 0.6704$ for conversion from Table~\ref{tab:LOFAR-data}) at $z = 9.1$. The white dashed line shows the maximum power of the models in the standard case (the corresponding likelihood value is $\l=0.4898$). Magenta data points correspond to the LOFAR data from Table~\ref{tab:LOFAR-data} (two-sided error bars). Right: corresponding thermal histories, i.e., evolution of the mean temperature of neutral intergalactic gas with redshift. Each curve is shown down to the (model-dependent) redshift of end of reionization.}
\label{Fig:modelsFialkov2}
\end{figure*}

%We have applied our likelihood framework using the limits on the 21-cm power spectrum derived from 141 hours of LOFAR observations (Table~\ref{tab:LOFAR-data}). 
As an illustration,  in Figure~\ref{Fig:modelsFialkov2} we show the complete set of excess-background power spectra (7702 models in total) colour-coded by the probability that the data is consistent with the model. For comparison, we also show  the maximum power of the models in the standard case (white line). The upper limits from \citet{LOFAR-EoR:2020} are plotted for reference. As we see from the figure, the current observational limits from LOFAR are strong enough to rule out a significant fraction of the explored excess-background scenarios (all corresponding to a cold IGM with $\sim 50\%$ ionization at $z=9.1$, as we will see later). However, for the standard astrophysical scenarios where the values of the power spectra are lower, only the most extreme models can be ruled out, and only in the lowest $k$-bin. A set of corresponding thermal histories is plotted in the right panel of Figure~\ref{Fig:modelsFialkov2}. The LOFAR upper limits by \citet{LOFAR-EoR:2020} disfavour a late X-ray heating which leaves the IGM cold for most of the EoR. Scenarios with early X-ray heating cannot be ruled out by the data as, typically, the corresponding power spectrum values are low.

\section{Results}
\label{sec:Res_radio}
Using the predicted values of the  spherically averaged binned power spectrum in all seven $k$-bins we can rule out scenarios which yield strong fluctuations at $z=9.1$. In the standard scenario with the CMB as a  background radiation, a few factors need to come together to ensure maximum power. First, the spin temperature has to be  fully coupled to the gas temperature, which, for realistic star formation scenarios, is guaranteed to be the case at $z= 9.1$ \citep[e.g.,][]{Cohen:2018}. Second, the larger the contrast between $T_{\rm gas}$ and $T_{\rm rad}$, the stronger the signal. For $T_{\rm rad} = T_{\rm CMB}$, the strongest contrast between the two temperatures is reached in cases of cold IGM. In the case of the  excess-background models, the coupling is less efficient compared to the standard models, however  the signals are enhanced  due to the larger contrast between the gas temperature and the temperature of the background radiation. Similarly to the standard case, the deepest signals correspond to the scenarios with the inefficient X-ray heating. Finally, fluctuations in the gas temperature and the neutral fraction play a role. Because here we have chosen a hard X-ray spectrum \citep{fragos13, Fialkov2014}, heating is nearly homogeneous, and the dominant source of fluctuations at $z=9.1$ is the non-uniform process of reionization with peak power at $\sim50\%$ ionization fraction. For a fixed thermal history, nearly homogeneous reionization would result in a smoother signal and, thus, lower power of the 21-cm fluctuations, compared to a patchy reionization scenario.

\subsection{Limits on the Excess-Radio Background}
\begin{figure*}
\centering
\psfrag{f*}[c][c][1.2]{$f_*$}
\psfrag{Vc}[c][c][1.2]{$V_c$ [km s$^{-1}$]}
\psfrag{fx}[c][c][1.2]{$f_X$}
\psfrag{tau}[c][c][1.2]{$\tau$}
\psfrag{Rmfp}[c][c][1.2]{$R_{\rm mfp}$ [Mpc]}
\psfrag{A}[c][c][1.2]{$A_{\rm r}$}
\includegraphics[width=1.0\textwidth, angle=0]{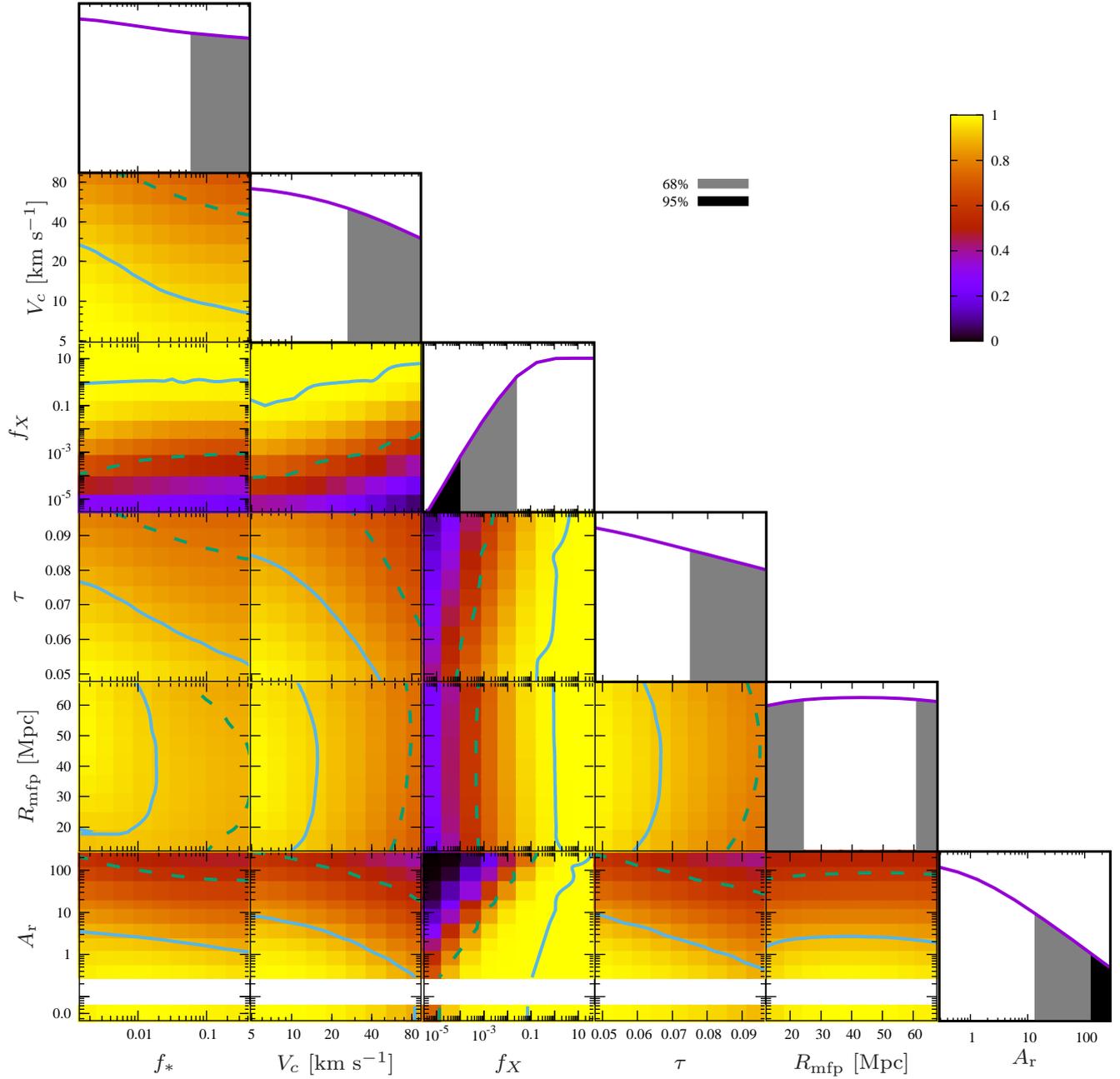}
\caption{1D and 2D marginalised likelihood of the astrophysical parameters ($f_*,~V_c,~f_X,~\tau,~R_{\rm mfp}, ~A_{r}$) obtained using excess-background models. In addition, we append the normalized likelihood values for our standard models below the white band of the bottom row to highlight the consistency with the excess-background case. The standard-case normalized likelihood was calculated by using a joined set of the excess-background models and standard models. Regions of 2D marginalised likelihoods which are on the darker side (red, purple and black)  of the solid lines are disfavoured with more than 39\% (1-$\sigma$ in 2D) confidence, and the regions which are on the darker side of the dashed lines are disfavoured with more than 86\% (2-$\sigma$ in 2D) confidence. The grey regions in the 1D likelihood distribution are also disfavoured at the 68\% confidence level, and the black regions are disfavoured at the 95\% confidence level. We do not show the 99\% C.L. for $A_{\rm r}$ here as it is close to the upper limits on the prior (set by the LWA1 limit). All limits are listed in Table \ref{tab:results}.}
\label{fig:2d_radio}
\end{figure*}

Using $\l(\vec\theta)$, we calculate the normalized probability for each of the parameters, $\vec\theta= [f_*,~V_c,~f_X,~\tau,~R_{\rm mfp}$, $A_{\rm r}]$, and parameter pairs, marginalising over the rest of the parameter space. The resulting probability distributions are normalized using the criterion that the total probability (area under the curve) is 1 within the considered prior ranges. The resulting two-dimensional and one-dimensional probabilities of all the model parameters are shown in Figure~\ref{fig:2d_radio}, where we  divided each probability function by its peak value to show the marginalised likelihood of all possible combinations uniformly. Using one-dimensional probabilities we find the 68\%, 95\% and 99\% confidence intervals for $A_{\rm r}$, and 68\% and 95\% confidence intervals for $f_X$, while the constraints on the other parameters are weaker and could be inferred only at 68\% level (see  Table~\ref{tab:results}). We calculate  each confidence level  (C.L.) by selecting parameter-bins with the highest probability up to the corresponding  cumulative probability (e.g.,  of 0.68 for the 68\% C.L.). We also note the limits  where the one-dimensional probabilities are below $\exp(-1/2)$ of the peak  (similar to the Gaussian 1-sigma definition).

Marginalising over the residual model parameters ($f_*,~V_c,~f_X,~\tau,~R_{\rm mfp}$), we derive constraints on $A_{\rm r}$ finding that  LOFAR upper limit rules out $A_{\rm r} > 15.9$ at 68\%, $A_{\rm r} > 182$ at 95\% and $A_{\rm r} > 259$ at 99\%, equivalent to 0.8\%, 9.6\% and 13.6\% respectively of the CMB at 1.42 GHz. The 95\% limit on $A_{\rm r}$ of 182 is equivalent to 262 mK at 1.42 GHz and is within 3$\sigma$ of the LWA1 measurement. The likelihood, which peaks at low values of $A_{\rm r}$,  drops by a factor of $\exp(-1/2)$ by $A_{\rm r} = 60.9$ corresponding to 3.2\% of the CMB at 1.42 GHz. In our analysis we have fixed the value of the spectral index of the radio background to $\beta = -2.6$. We have checked that the uncertainty in the spectral index $\Delta \beta = 0.05$, reported by LWA1 \citep{Dowell:2018}, would lead to only up to $\sim 3$ per cent variation in the intensity of the excess radio background at the frequency of 140 MHz corresponding to $z = 9.1$. 
 
\begin{table*}
\begin{center}
\caption{Limits on astrophysical parameters and the derived IGM parameters. From left to right: type of model and constraint; mean temperature of neutral gas at $z=9.1$ in K; ionization fraction of the IGM; duration of reionization defined as the redshift interval between 90\% neutral IGM and 10\% neutral; redshift of the mid-point of reionization (defined as the redshift at which neutral fraction is 50\%); star formation efficiency; minimum circular velocity of star forming halos in km\,s$^{-1}$; X-ray heating efficiency; CMB optical depth; mean free path of ionizing photons in comoving Mpc; amplitude of the excess radio background compared to the CMB at the  reference frequency of 78 MHz (as defined in Eq.~\ref{Eq:Trad}). For the case of excess radio background (Ex. bck. in the table) we show both 68\% limits (top row), 95\% limits (second row) and 99\% limits (third row). We also find the parameter values at which the likelihood drops to $\exp(-1/2)$ of the peak value (third row). In the standard case we can only show the 68\% limits, as the 1D PDFs are rather flat.}
\label{tab:results}
\begin{tabular}{|l|l|c|c|c||c|c|l|c|c|c|}
\hline
Model &  $T_{\rm gas}$ & $\xb$ & $\Delta z$ & $z_{\rm re}$ & $f_*$ & $V_c$ & $f_X$ & $\tau$ & $R_{\rm mfp}$ & $A_{\rm r}$ \\
\hline
Ex. bck., 68\% &  $> 16.1$ &  $< 38\%$ or $> 72\%$ & $> 3$ & $< 8.21$ & $< 0.05$  & $< 28$ & $> 1\times10^{-2}$& $< 0.076$ & $>24$ and $< 60$ & $< 15.9$ \\
Ex. bck., 95\% &  $>2.89 $ &  NA & NA & NA & NA & NA & $>1 \times 10^{-4}$ & NA & NA & $<182$ \\
Ex. bck., 99\% &  $>2.25 $ &  NA & NA & NA & NA & NA & NA & NA & NA & $<259 $ \\
Ex. bck., $e^{-1/2}$ &  $> 6.0$ & NA & NA & NA & NA  & NA & $> 8\times 10^{-4}$ & NA & NA & $< 60.9$ \\
\hline
Standard, 68\% &  $> 10.1$ & $< 38\%$ or $> 72\%$ & $> 3$ & $< 8.51$ & $< 0.05$ & $< 36$ & $> 5\times 10^{-3}$ & $< 0.080$ & $<30$ or $>49$& NA \\
\hline
\end{tabular}
\end{center}
\end{table*}

%\begin{table*}
%\begin{center}
%\caption{limits on astrophysical parameters and the derived IGM parameters obtained using excess-background models. We have derived these limits from the regions where the one-dimensional probabilities are below $\exp(-1/2)$ of the peak (similar to the Gaussian 1-sigma definition).}
%\label{tab:results1}
%\begin{tabular}{|c|c|c|c||c|c|c|c|c|c|}
%\hline
%$T_{\rm gas}$ & $\xb$ & $\Delta z$ & $z_{\rm re}$ & $f_*$ & $V_c$ & $f_X$ & $\tau$ & $R_{\rm mfp}$ & $A_{\rm r}$ \\
%\hline
%$> 6$ & NA & NA & NA & NA  & NA & $> 8\times 10^{-4}$ & NA & NA & $< 60.9$ \\
%\hline
%\end{tabular}
%\end{center}
%\end{table*}

\citet{Fialkov:2019} showed that the global signal reported by EDGES Low-Band can be produced by adding an extra radio background with $1.9 < A_{\rm r} < 418$ relative to the CMB at the 78\,MHz reference frequency (corresponding to $0.1 - 22$ per cent of the CMB at 1.42\,GHz). Even though part of this range is now ruled out by the new LOFAR limits, models with values of $A_{\rm r}$ between 0.1\% and 9.6\% (at 95\% C.L.) of the CMB at 1.42 GHz are still allowed and could fit the EDGES Low-Band detection. Such a small contribution is within the measurement error of LWA1 \citep[][report excess background of $603^{+102}_{-92}$\,mK at the 21-cm rest frame frequency of 1.42 GHz]{Dowell:2018} and would remain a plausible explanation for the detected EDGES signal even if the  excess measured by ARCADE2 and LWA1 is due to an erroneous Galactic modelling \citep{Subrahmanyan:2013}. 
%This limit, for the first time, rules out strong contribution of the high-redshift Universe to the excess detected by ARCADE2 and LWA1.

\begin{figure}
\includegraphics[width=3.2in]{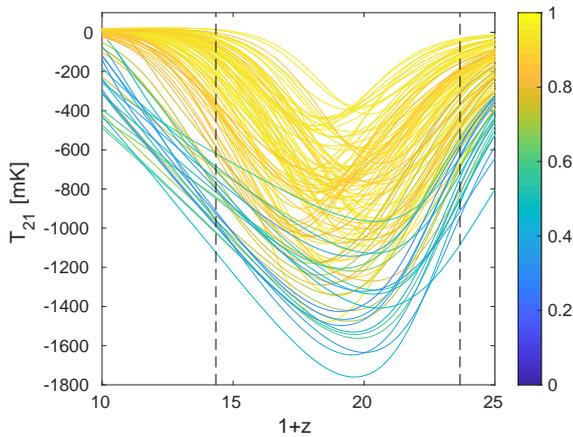}
\caption{Excess-background global signals that are consistent with EDGES Low-Band, colour-coded with respect to their model likelihood values under LOFAR (Eq.~\ref{eq:like1}) as is indicated  on the colour bar. Vertical dashed lines mark the frequency range $60-99$ MHz over which the best-fit detected signal was reported \citep{Bowman:2018}.}
\label{Fig:globalEDGES}
\end{figure}

In Figure~\ref{Fig:globalEDGES}, as an illustration, we show global 21-cm signals for those excess-background models from our data set that are broadly consistent with the tentative EDGES Low-Band detection. In order to define this consistency we follow the simple approach taken by \citet{Fialkov:2019} by requiring the signal  to be deep, and localised within the band of the EDGES Low instrument. Within $99\%$ confidence, the cosmological signal should satisfy 
\begin{eqnarray}
300\rm mK < && \bigl\{ \rm max \left[T_{21}(60<\nu<68)\right] \nonumber \\
&&  -\rm min\left[T_{21}(68<\nu<88)\right]\bigr\}  < 1~\textrm{K},
\label{Eq:C2}
\end{eqnarray}
and 
\begin{eqnarray}
300\rm mK < && \bigl\{ \rm max\left[T_{21}(88<\nu<96)\right] \nonumber \\
&& -\rm min\left[T_{21}(68<\nu<88)\right]\bigr\}  < 1~\textrm{K}.
\label{Eq:C3}
\end{eqnarray}
The signals in Figure~\ref{Fig:globalEDGES} are colour-coded with respect to the LOFAR likelihood (same as in Figure~\ref{Fig:modelsFialkov2}). All the signals consistent with EDGES Low-Band have relatively high LOFAR likelihood, $\l\geq0.31$. This is because the EDGES detection implies an early onset of the Ly-$\alpha$ coupling \citep{Schauer:2019} due to efficient star formation ($f_*> 2.8$\%) in lower-mass halos with circular velocity below $V_c = 45\,$km\,s$^{-1}$   \citep[corresponding to $M_h < 7.8 \times 10^8\,$M$_{\sun}$ at $z = 17$,][]{Fialkov:2019}. In such models the IGM is heated and partially ionized by $z=9.1$, resulting in relatively low-intensity 21-cm signals in the LOFAR band.

\subsection{Astrophysical Limits}
\begin{figure*}
\centering
\psfrag{f*}[c][c][1.2]{$f_*$}
\psfrag{Vc}[c][c][1.2]{$V_c$ [km s$^{-1}$]}
\psfrag{fx}[c][c][1.2]{$f_X$}
\psfrag{tau}[c][c][1.2]{$\tau$}
\psfrag{Rmfp}[c][c][1.2]{$R_{\rm mfp}$ [Mpc]}
\psfrag{A}[c][c][1.2]{$A_{\rm r}$}
\includegraphics[width=1.0\textwidth, angle=0]{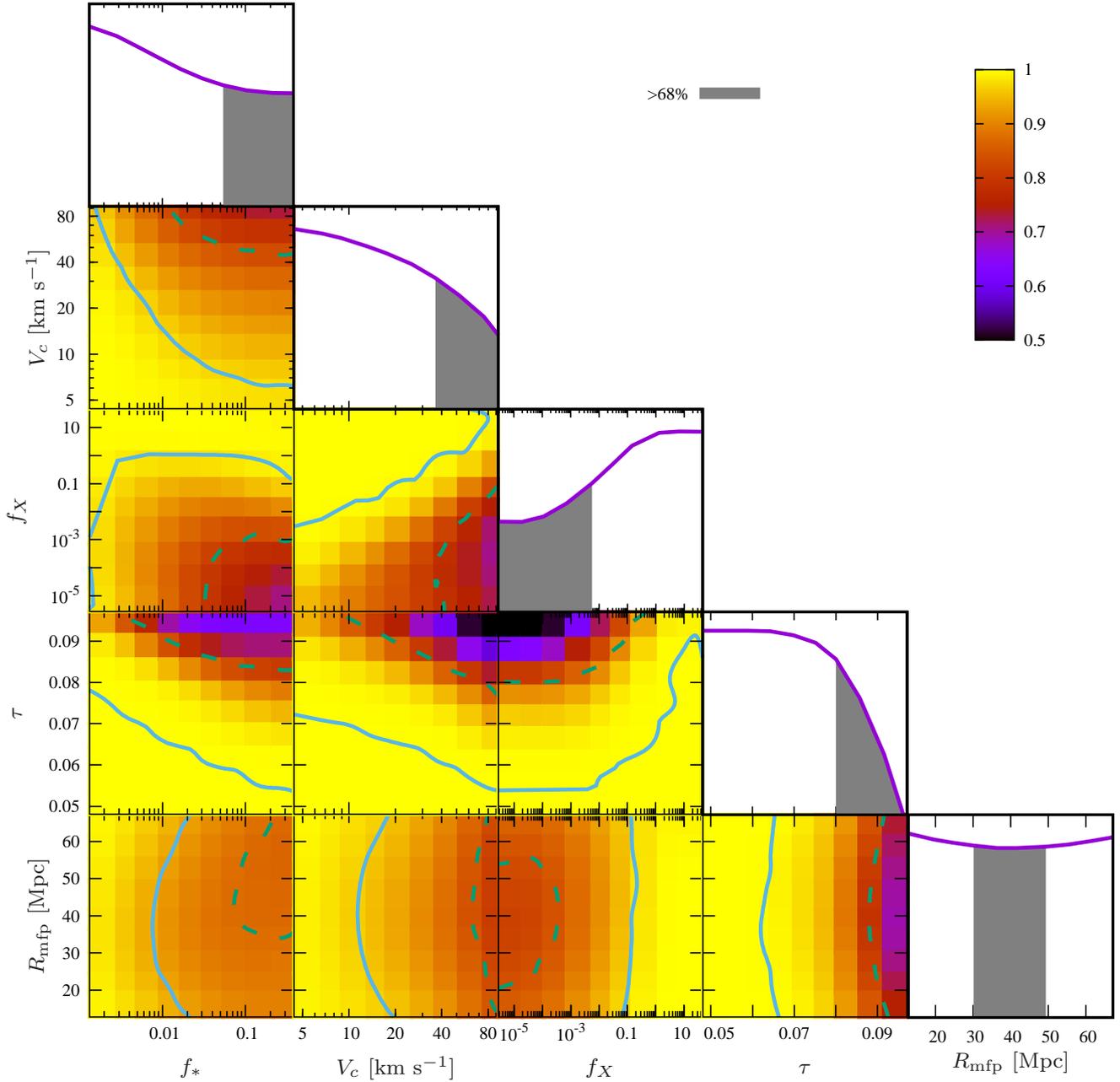}
\caption{1D and 2D marginalised likelihood of the astrophysical parameters ($f_*,~V_c,~f_X,~\tau,~R_{\rm mfp}$) obtained using standard models ($A_{\rm r} =0$). The regions of 2D marginalised likelihoods which are on the darker side of the solid lines are disfavoured with more than 39\% (1-$\sigma$ in 2D) confidence, and the regions which are on the darker side of the dashed lines are disfavoured with more than 86\% (2-$\sigma$ in 2D) confidence. The grey regions in the 1D likelihood distribution are also disfavoured at the 68\% confidence level (also listed in Table \ref{tab:results}). Note that the colour scale is not the same as that in Figure~\ref{fig:2d_radio}.}
\label{fig:2d}
\end{figure*}

Next, we explore the implications of the LOFAR upper limits for the rest of the model parameters ($f_*,\,V_c,\,f_X,\,\tau,\,R_{\rm mfp}$). In this work we assume hierarchical structure formation with a simple prescription for the formation of stars and X-ray binaries. Therefore, LOFAR limits at $z=9.1$ can be used to constrain properties of the first star forming halos (appearing at $z\sim 30-60$ in our simulations) and first sources of light at Cosmic Dawn. The resulting two-dimensional and one-dimensional probabilities are shown in Figure~\ref{fig:2d_radio} and the limits are summarised in Table~\ref{tab:results}. In the limiting case of the negligible radio background our results converge to the standard case with the CMB as the background radiation. This trend is demonstrated in Figure~\ref{fig:2d_radio} where the two-dimensional probabilities of standard models, with $A_{\rm r}=0$, are appended below the white band. For completeness we also explore the set of standard models separately, showing their two-dimensional and one-dimensional probabilities in Figure~\ref{fig:2d} and listing the corresponding 68\% limits in Table~\ref{tab:results}.

All disfavoured models feature efficient star formation with $f_* \ga 5\%$  at 68\% C.L. (Table~\ref{tab:results}). However, the corresponding 1D marginalised likelihood is rather flat and never drops below a factor of $\exp(-1/2)$ relatively to its peak value. Higher values of $f_*$ result in stronger fluctuations which are easier to rule out. Higher values of $f_*$ also imply  stronger Ly-$\alpha$ background and, thus, an  earlier onset of Ly-$\alpha$ coupling which yields signals with larger amplitudes  \citep[e.g.,][]{Cohen:2019}.

Another model parameter related to star formation in first halos is $V_c$. Higher $V_c$  is equivalent to larger minimum mass of star forming halos which are more strongly clustered, thus yielding stronger fluctuations. In the hierarchical model of star formation that we adopt here, higher  $V_c$ also implies later onset of star formation and X-ray heating. In such models chances are that fluctuations (e.g., heating) are not yet saturated by $z=9.1$ resulting in stronger 21-cm signals that can be ruled out by LOFAR. We find that values of $V_c$ above 28\,km\,s$^{-1}$ (corresponding to $4.5\times 10^8\,$M$_{\sun}$ at $z=9.1$) are disfavoured by the data at 68\% (the corresponding 1D marginalised likelihood is rather flat and never drops below the threshold value of $\exp(-1/2)$ relatively to its peak value). The standard-physics limit is 36\,km\,s$^{-1}$, or $9.5\times 10^8\,$M$_{\sun}$ at $z=9.1$. Even though the limits on $V_c$ are weak at the moment, the LOFAR data favour the  existence of low-mass halos  \citep[in agreement with EDGES High Band results,][]{Monsalve:2019}.

In our models gas temperature is regulated by the interplay between several cooling and heating mechanisms with the major roles played by adiabatic cooling due to the expansion of the universe and X-ray heating by X-ray binaries, although the latter is partially degenerate with $f_*$ and $V_c$ which regulate the number of X-ray binaries\footnote{The degeneracy is visible in the 2D PDFs of $f_*-f_X$  and $V_c-f_X$ shown in Figures~\ref{fig:2d_radio} and \ref{fig:2d}.}. Therefore, the X-ray efficiency of the first X-ray binaries is directly constrained by LOFAR with a values $f_X<1\times 10^{-2}$ disfavoured at 68\% C.L. and $f_X<1\times 10^{-4}$ disfavoured at 95\% C.L., implying a lower limit on the total X-ray luminosity per star formation rate (Eq.~\ref{Eq:fX}) of $3\times 10^{38}\,{\rm erg\,s}^{-1}M_{\sun}^{-1}\,{\rm yr}$ and  $3\times 10^{36}\,{\rm erg\,s}^{-1}M_{\sun}^{-1}\,{\rm yr}$ respectively. The 1D likelihood, which peaks at high $f_X$ values, is steep enough and drops below the threshold $\exp(-1/2)$ of its peak value at $f_X = 8\times 10^{-4}$ (corresponding to $2.4\times 10^{37}\,{\rm erg\,s}^{-1}M_{\sun}^{-1}\,{\rm yr}$). 
In the standard case only the 68\% limit can be calculated and is $f_X<5\times 10^{-3}$ ($1.5\times 10^{38}\,{\rm erg\,s}^{-1}M_{\sun}^{-1}\,{\rm yr}$ respectively).

The current LOFAR data also disfavour models with mid-point of reionization at $z\sim 9$. In such models the peak-power from ionizing fluctuations  falls within the currently-analysed LOFAR band, and, consequently, such models are relatively easy to exclude.  This constraint can be mapped on to limits on $\tau$: scenarios with $\tau > 0.076$ (excess background) or $\tau>0.080$ (standard models) are disfavoured at 68\%.  In both theories, the 1D likelihood curves of $\tau$ peak at low values of $\tau$ but do not drop below the threshold value of $\exp(-1/2)$ within the prior ranges. Finally, we find that the constraints on the model parameter $R_{\rm mfp}$ are very weak, with the 1D marginalised likelihood being very flat. This means that our model power spectrum is not sensitive to the changes in $R_{\rm mfp}$ value at $z \sim 9$.
 
\subsubsection{Comparison with EDGES}
Focusing on the standard models we can compare the LOFAR limits reported above to the limits extracted from the data of the global 21-cm instrument EDGES High-Band ($90-190$ MHz, correspond to the 21-cm signal from $z=6-15$). Using a similar set of standard models and similar prior ranges of parameters as we explore here, \citet{Monsalve:2019} found that the EDGES High-Band data favour (at 68\% confidence) the following parameter ranges assuming a fixed X-ray SED \citep[softer than what we use here; however, the global signal constraints prove to be nearly insensitive to the X-ray SED,][]{Monsalve:2019}: $R_{\rm mfp} <36.1$ Mpc, $V_c<21.5$\,km\,s$^{-1}$ (equivalent to $2\times 10^8\,$M$_{\sun}$ at $z=9.1$), $f_X>2.5\times 10^{-3}$, $f_*<0.4\%$ or $f_*>3.6\%$ (signals with both lower and higher values of $f_*$ are likely to be outside of the band of EDGES High),  $\tau <0.072$ or $0.074<\tau< 0.079$ (where the second band is most likely due to the instrumental systematic). Overall LOFAR and the EDGES High-Band experiment are in agreement ruling out scenarios with inefficient X-ray heating and models in which the Universe was ionized by massive halos only (of mass few$\,\times 10^8\,$M$_{\sun}$ or higher, at $z\sim 9.1$). Similar trends  were found with the SARAS2 data \citep[although only 264 models were examined in that case,][]{Singh:2017}. 

\subsection{Limits on  the Thermal and Reionization Histories}
\begin{figure*}
\centering
\psfrag{DeltaZ}[c][c][0.99]{$\Delta z$}
\psfrag{Zre}[c][c][0.99]{$z_{\rm re}$}
\psfrag{Tgas}[c][c][0.99]{$T_{\rm gas}$ [K]}
\psfrag{xHII}[c][c][0.99]{$\xb$}
\includegraphics[width=0.48\textwidth, angle=0]{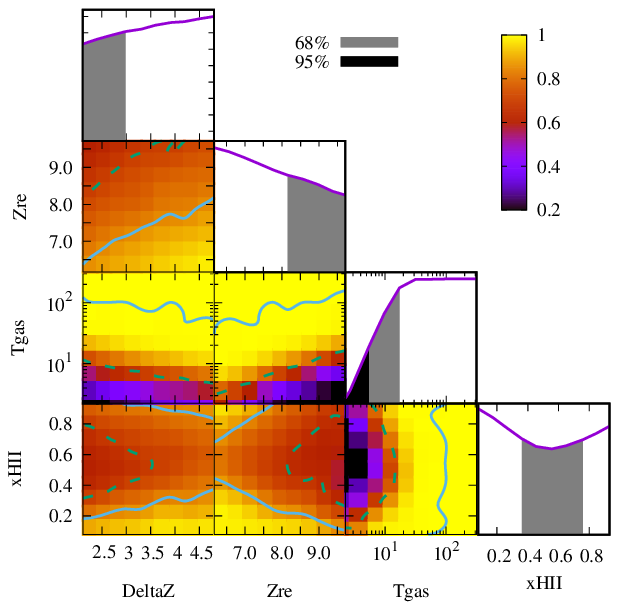}
\hspace{0.02\textwidth}
\includegraphics[width=0.48\textwidth, angle=0]{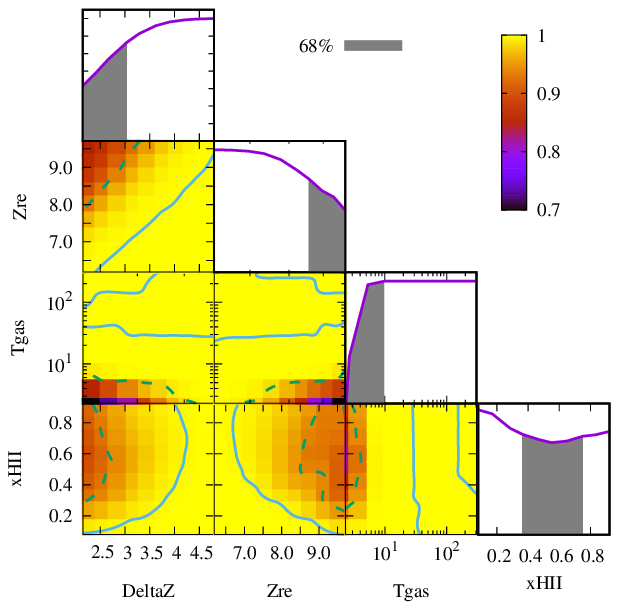}
\caption{1D and 2D marginalised likelihood of the IGM parameters [$T_{\rm gas}, \xb, \Delta z, z_{re}$] obtained using excess-background models (left) and standard models (right). Note separate colour bars (top right of each panel). The regions of 2D marginalised likelihoods which are on the darker side of the solid lines are disfavoured with more than 39\% (1-$\sigma$ in 2D) confidence, and the region which is on the darker side of the dashed line is disfavoured with more than 86\% (2-$\sigma$ in 2D) confidence. The grey regions in the 1D likelihood distribution are also disfavoured at the 68\% confidence level, and (for excess-background models) the black region is disfavoured with more than 95\% C.L. The limits are listed in Table \ref{tab:results}. Note that the colour scales are not the same as those in Figures~\ref{fig:2d_radio} and \ref{fig:2d}.}
\label{fig:2d_global_radio}
\end{figure*}

We use the LOFAR upper limits on the 21-cm power spectrum to put limits on the thermal and ionization state of the IGM at $z=9.1$. We repeat the likelihood calculation applying it to the IGM parameters $\vec\theta= [T_{\rm gas},\,\xb,\,z_{\rm re},\,\Delta z]$. The resulting two-dimensional and one-dimensional probabilities of $T_{\rm gas}$, $\xb$, $z_{\rm re}$ and $\Delta z$ are shown in Figure~\ref{fig:2d_global_radio} (left panel shows the case of the extra radio background, while the standard case is shown on the right for comparison). Our results are also summarised in Table~\ref{tab:results}. We have also tabulated the limits obtained from the regions where the one-dimensional probabilities are below $\exp(-1/2)$ of the peak.

As we see from the figure and the table, the LOFAR data indeed disfavour scenarios with cold IGM. The lower limit on the temperature of neutral gas at $z=9.1$ is $16.1\,$K at 68\% C. L. (while it is only $10.1\,$K in the standard case) $2.89\,$K at 95\% C. L. and $2.25\,$K at 99\% C. L. The likelihood, which peaks at high values of $T_{\rm gas}$, drops by a factor of $\exp(-1/2)$ at $T_{\rm gas} = 6$\,K in the excess-background case. As expected, there is some degree of degeneracy between the constraints on the thermal and reionization histories with the strongest limits on temperature coming from the cases with mid-point of reionization occurring at $z\sim 9$.

Through marginalising over the thermal histories we can put limits on the process of reionization (Figure~\ref{fig:2d_global_radio} and Table~\ref{tab:results}). We find that the LOFAR limits disfavour fast reionization scenarios (with $\Delta z\lesssim 3$)  with ionized fractions between $\sim 38\%$ and $\sim 72$\%  at $z=9.1$. The high end of the allowed $\xb$ values ($\xb>72\%$ at $z=9.1$) is inconsistent with other probes of reionization and would be ruled out if joined constraints were considered: e.g., Ly-$\alpha$ damping wing absorption  in the spectrum of the quasar at $z=7.54$  suggests that the Universe is $\sim 60\%$ neutral at that redshift \citep[ionization fraction less than 40\%,][]{Banados:2018, Davies:2018}. The quantitative joint analysis, however, is beyond the scope of this paper.

\subsection{The optimal exclusion space}
In the analysis above we considered  two separate data sets: the model parameters with $\vec\theta = [f_*,\,V_c,\,f_X,\,\tau,\,R_{\rm mfp},\,A_{\rm r}]$ and the derived IGM parameters with $\vec\theta= [T_{\rm gas},\,\xb,\,z_{\rm re},\,\Delta z]$. We showed that LOFAR is most sensitive to the radio background amplitude and to the thermal history of the IGM. To strengthen this point here we focus our attention on just these two parameters, i.e.,  $A_{\rm r}$ and $T_{\rm gas}$. These parameters are independent as the gas temperature is determined by properties of astrophysical sources (mainly $V_c$, $f_*$ and $f_X$), while $A_{\rm r}$ is the amplitude of the phenomenological radio background.

%The entire analysis until now has provided constraints on the parameters at the 68\% confidence level. This is due to two reasons (1) current LOFAR power spectrum upper limits are too high, and (2) 95\% (2$\sigma$) limits are not achievable within prior ranges considered here for most of the parameters. However, it is particularly important to explore the implications of the LOFAR upper limits for $A_{\rm r}$ and $T_{\rm gas}$ more thoroughly. As explained above, our model crucially depends on these two parameters and the LOFAR limits can be used to constrain them at 95\% C.L. 

\begin{figure}
\centering
\psfrag{Ar}[c][c][0.99]{$A_{\rm r}$}
\psfrag{Tgas}[c][c][0.99]{$T_{\rm gas}$ [K]}
\includegraphics[width=0.48\textwidth, angle=0]{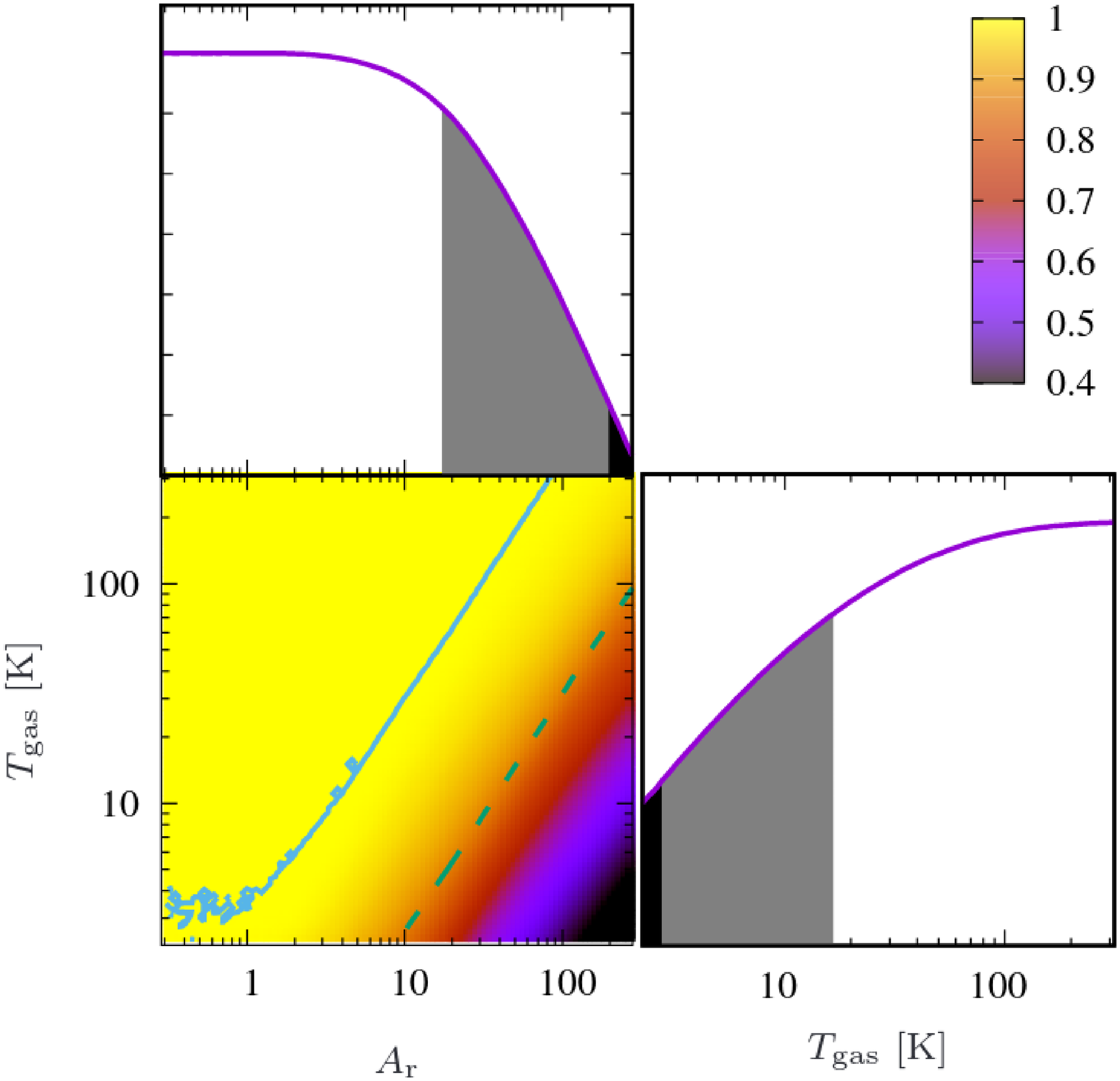}
\caption{The optimal exclusion space of LOFAR: 1D and 2D marginalised likelihood of the excess-radio background parameter $A_{\rm r}$ and the IGM temperature $T_{\rm gas}$ obtained using excess-background models. The region of 2D marginalised likelihood which is on the darker side of the solid lines is disfavoured with more than 39\% (1-$\sigma$ in 2D) confidence, and the region which is on the darker side of the dashed line is disfavoured with more than 86\% (2-$\sigma$ in 2D) confidence. The grey regions in the 1D likelihood distributions are also disfavoured at the 68\% confidence level, and the black regions are disfavoured at the 95\% confidence level. To produce this figure we have marginalised over $V_c$, $f_X$ and $f_*$ and assumed fixed values of $R_{\rm mfp} = 40$ comoving Mpc and $\tau = 0.055$.}
\label{fig:Ar-tgas}
\end{figure}

We calculate the normalized probability for $A_{\rm r}$ and $T_{\rm gas}$ marginalising over $V_c$, $f_*$ and $f_X$. For simplicity we fix the reionization parameters $R_{\rm mfp} = 40$ comoving Mpc and $\tau = 0.055$. We show the resulting 2D and 1D probabilities  in Figure~\ref{fig:Ar-tgas}. As a sanity check, we calculate  68\% (grey), 95\% (black) and 99\% (not shown) confidence intervals finding that, at 95\% confidence level, LOFAR data rule out  $A_{\rm r} > 182$ and $T_{\rm gas}<2.89$ K at $z=9.1$ and at 99\%  LOFAR data rule out  $A_{\rm r} > 259$ and $T_{\rm gas}<2.25$ K. These limits are consistent with our previous numbers in Table~\ref{tab:results}. 

\section{Qualitative Comparison with Previous Results}
\label{Sec:compare}
\citet{ghara2020} explored the implications of the LOFAR data in terms of the astrophysical parameter and statistical constraints on the IGM properties assuming standard-physics  models (with the CMB as the background radiation). We verify the consistency of our conclusions with \citet{ghara2020} by qualitatively comparing our standard case results for the thermal and ionization states of the IGM. A quantitative  comparison between the two works is not possible at this stage because of the different choices of modelling, parameterization and priors. Moreover, because the 21-cm signal is sensitive to the thermal and ionization histories  the values of the gas temperature and ionization fraction can be directly constrained using the data. However, the mapping between these quantities and the astrophysical properties of the UV and X-ray sources (in our case $f_*,~V_c,~f_X,~\tau$ and $R_{\rm mfp}$) is model-dependent. Therefore, in this paper we refrain from comparing the astrophysical constraints leaving it for future work.

In their work \citet{ghara2020} explored two scenarios: (1) homogeneous spin temperature, which implies saturated Ly-$\alpha$ background and homogeneous X-ray heating. The parameters that are varied in this case include gas temperature (or, equivalently, spin temperature), minimum halo mass and ionizing efficiency.  (2) Inhomogeneous heating by soft X-ray sources with power-law SED where $M_{\rm min}$ and the spectral index of X-ray sources  were kept fixed; the parameters that were varied are ionizing efficiency, X-ray efficiency (defined differently than in our work) and minimum mass of X-ray emitting halos. In all cases the value of star formation efficiency was kept fixed at $f_*=2\%$. In comparison, we explore the popular case of heating by a realistic population of X-ray binaries with hard SED. In this case heating is inefficient and fluctuations are smoothed out \citep{Fialkov2014}. Therefore, we expect our results to be closer to case (1)  of \citet{ghara2020}. Moreover, in our work all the parameters (except for X-ray SED) are allowed to vary over a wide range, e.g., $f_*$ is varied between 0.1\% and 50\%. 

Despite these differences in  modelling, qualitatively our work is consistent with  \citet{ghara2020}. Both works rule out a cold IGM with an ionization fraction close to $50\%$ at $z = 9.1$. Namely, in their case (1) $\xb \sim 0.24-0.6$ and $T_{\rm gas}\lesssim 3$\,K are disfavoured (at 95\%); while we find that $\xb \sim 0.38-0.72$ and $T_{\rm gas}\la 10.1$\,K  are  disfavoured (at 68\%).

\section{Conclusions}
\label{sec:conc}

%LOFAR limits on reionization are still very weak in comparison, and constraints on the gas temperature remain the unique strength of the 21-cm experiments even in the case of the excess-background models.

%Using LOFAR upper limits based on 141 hours of observations and assuming existence of isotropic extra radio background in addition to the CMB, we have derived 68\% C.L. limits on its amplitude along with the limits on the astrophysical parameters of cosmic dawn and reionization. For completeness we also  have considered two classes of models:  excess-background model with phenomenological and standard astrophysical scenario. 

In this paper we have  used the upper limit on the 21-cm signal from $z=9.1$ based on 141 hours of observations with LOFAR \citep{LOFAR-EoR:2020} to evaluate the contribution of the high redshift Universe to the excess radio background over the CMB detected by ARCADE2  \citep{Fixsen:2011} and LWA1 \citep{Dowell:2018}. Assuming synchrotron spectrum of the radio background with spectral index $\beta = -2.6$ and marginalising over the astrophysical properties of star-forming sources, we find (at 95\% C.L.) the contribution above the CMB level to be less than a factor of 182 at the reference frequency of 78 MHz, equivalent to 9.6\% of the CMB at 1.42 GHz. This limit, for the first time, rules out strong contribution of the high-redshift Universe to the excess detected by ARCADE2 and LWA1. At the level below 9.6\% of the CMB, the extra radio background could, on one hand, be strong enough to explain the tentative EDGES Low-Band detection which requires an excess of at least $0.1\%$ of the CMB \citep{Fialkov:2019}. On the other hand, such a small contribution would  be within the measurement error (at 2$\sigma$ level) of the LWA1 radio telescope. Hence, it would remain a plausible explanation for the detected EDGES signal, even if the  excess radio background measured by ARCADE2 and LWA1 is due to an erroneous Galactic modelling \citep{Subrahmanyan:2013}.

We also use LOFAR data to constrain thermal and ionization state of the IGM at $z=9.1$ in models with and without the extra radio background over the CMB.  If such an extra radio background is present at z = 9.1, the fluctuations in the 21-cm signal are boosted compared to the standard case, which gives LOFAR a larger lever to reject models. Therefore, for the models with excess radio background, constraints on the astrophysical properties and the properties of the IGM are tighter than in the standard case. In particular, compared to the upper limit of 10.1\,K (at 68\% C.L.) in the standard case, warmer IGM scenarios with mean neutral gas temperature of up to 16.1\,K are disfavoured in the extra radio background models. In the latter case we were also able to derive 95\% and 99\% C.L. on temperature of 2.89\,K and 2.25\,K respectively. Thus, the LOFAR data rule out the cold IGM scenarios in which the gas is expected to have a temperature of 2.1 K at $z=9.1$ with 99.8\% C.L..

Using the LOFAR  data we have also derived 68\% C.L. limits on the astrophysical parameters of Cosmic Dawn and EoR. The data disfavour very efficient star formation above $5$\%, imply the existence of small halos at early times (of masses below few$\times10^8\,$M$_{\sun}$ at $z = 9.1$), require the presence of X-ray sources, and disfavour a CMB optical depth above $\tau \sim 0.076$. For the suite of standard models,  we point out that the LOFAR data rule out similar type of models as those rejected by the global signal experiments, namely  the EDGES High-Band \citep{Monsalve:2019} and SARAS2 \citep{Singh:2018}.   Finally, we note that our constraints of the standard-physics parameters  are in a qualitative agreement with the results reported  by \citet{ghara2020}.  A detailed comparison between these two works is beyond the scope of this paper.

Although other high redshift probes (e.g., the {\it Planck} measurement of the  CMB optical, high redshift quasars and galaxies) allow to put tighter constraints on the ionization history and properties of the UV sources at EoR, the 21-cm observations provide a unique way to probe the thermal history of the Universe  and test the nature of the radio background. Because quantities such as temperature and ionization fraction at the LOFAR redshift $z = 9.1$ are the results of cumulative (rather than an instantaneous) effect of star formation over the entire cosmic history, in this work we have refrained from using low-redshift constraints. Although the low-redshift observations constrain properties of bright galaxies during the EoR, they might be very different from the properties of high redshift sources owing to the redshift evolution of stellar population (e.g., as a result of the gradual process of metal enrichment).  

\section*{Acknowledgements}
This work was supported by the Science and Technology  Facilities Council [grant numbers ST/F002858/1 and ST/I000976/1] and the Southeast Physics Network~(SEPNet). We acknowledge the usage of the DiRAC HPC. We also acknowledge that the results in this paper have been achieved using the PRACE Research Infrastructure resources Curie based at  the Tr\`es Grand Centre de Calcul~(TGCC) operated by CEA near Paris, France and Marenostrum based in the Barcelona Supercomputing Center, Spain. Time on these resources was awarded by PRACE under PRACE4LOFAR grants 2012061089 and 2014102339 as well as under the Multi-scale Reionization grants 2014102281 and 2015122822. The authors gratefully acknowledge the Gauss Centre for Supercomputing e.V. (www.gauss-centre.eu) for funding this project by providing computing time through the John von Neumann Institute for Computing~(NIC) on the GCS Supercomputer JUWELS at J\"ulich Supercomputing Centre~(JSC). AF is supported by the Royal Society University Research Fellowship. Some of the numerical computations were done on the Apollo cluster at The University of Sussex. This project/publication was made possible for RB through the support of a grant from the John Templeton Foundation, as well as the ISF-NSFC joint research program (grant No.\ 2580/17). The opinions expressed in this publication are those of the authors and do not necessarily reflect the views of the John Templeton Foundation.

\section*{Data availability}
The data underlying this article will be shared on a reasonable request to the corresponding author.

\bibliographystyle{mnras} 
\bibliography{refs}

%--------------------------------------------------------------------

\appendix

\section{Artificial Neural Networks~(ANN)}
\label{ann}

\begin{figure*}
\centering
\psfrag{pk-ann}[c][c][1.2]{$P_{\rm ann}(k)$}
\psfrag{pk-true}[c][c][1.2]{$P_{\rm true}(k)$}
\psfrag{error}[c][c][1.2]{Error}
\psfrag{k=0.075}[c][c][1.2]{\hspace{1.3cm}$k=0.075\,h$Mpc$^{-1}$}
\psfrag{k=0.100}[c][c][1.2]{\hspace{1.3cm}$k=0.1\,h$Mpc$^{-1}$}
\psfrag{k=0.133}[c][c][1.2]{\hspace{1.3cm}$k=0.133\,h$Mpc$^{-1}$}
\psfrag{k=0.179}[c][c][1.2]{\hspace{1.3cm}$k=0.179\,h$Mpc$^{-1}$}
\includegraphics[width=1.0\textwidth, angle=0]{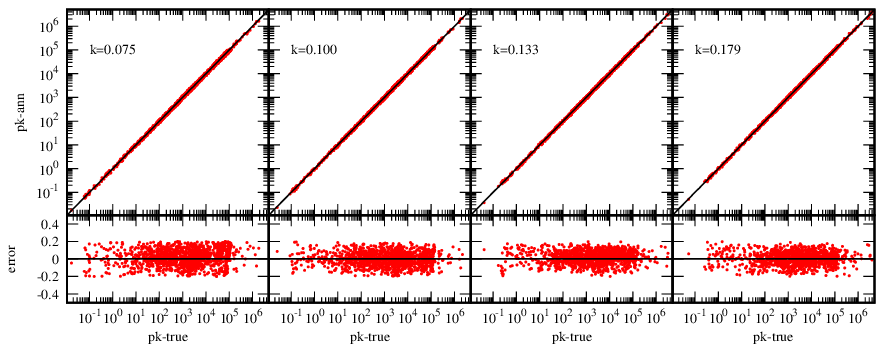}
\caption{The binned power spectrum values computed by the ANN emulator against the true values used for validation for the excess-background models. We show results only for four $k$-bins mentioned in the figure. The bottom panels show the relative difference. Note that the rms error (Table~\ref{tab:ann}) is calculated using data from all the seven $k$-bins.}
\label{fig:ANNerror}
\end{figure*}

\begin{table*}
\begin{center}
\caption{Specifications of different emulators used in our analysis. }
\label{tab:ann}
\begin{tabular}{cccccc}
\hline
Model & Number neurons & Parameters used & Number neurons & rms error & Result\\
& in the input layer & in the input layers & in the output layer &  $\sigma_{\rm ann}$ & \\
\hline
Excess-background & 6 & $[f_*,\,V_c,\,f_X,\,\tau,\,R_{\rm mfp},\,A_{\rm r}]$  & 14 & 7.2\% & Figure~\ref{fig:2d_radio} \\
Standard & 5 & $[f_*,\,V_c,\,f_X,\,\tau,\,R_{\rm mfp}]$ & 14 & 6.4\% & Figure~\ref{fig:2d}\\
Excess-background & 4 &  $[T_{\rm gas},\,\xb,\,z_{\rm re},\,\Delta z]$  & 14 & 5\% & Figure~\ref{fig:2d_global_radio} (left)\\
Standard & 4 & $[T_{\rm gas},\,\xb,\,z_{\rm re},\,\Delta z]$ & 14 &  4\% & Figure~\ref{fig:2d_global_radio} (right)\\
\hline
\end{tabular}
\end{center}
\end{table*}

Both numerical and semi-numerical simulations are too slow and computationally expensive to sample the 21-cm signal parameter space effectively. \citet{Jennings2019} evaluated the performance of five machine learning algorithms, and found deep Artificial Neural Networks~(ANN) to be the most efficient and best performing model to predict the 21-cm power spectra.

In order to produce a data set large enough to perform the statistical analysis outlined in Section~\ref{sec:method}, we use an emulator for the 21-cm power spectra based on an ANN \citep[see][for more details on methodology]{Fling:2019}. This allows us to sample the parameter space more thoroughly than the simulations permit and produce the large ensemble of models necessary for our analysis. The network was trained on the semi-numerical simulations for both the excess-background and standard cases described in Section~\ref{sec:sim_setup}, then used to predict the binned spherically averaged power spectra and sample variances (Section~\ref{sec:bias}) at $z  = 9.1$ in all seven LOFAR $k$-bins (see Table~\ref{tab:LOFAR-data}).

Our neural network was built with the python package Keras\footnote{\url{https://keras.io}}, which runs on top of Tensorflow\footnote{\url{https://www.tensorflow.org}}. The network consists of four hidden layers of sizes, 300, 300, 300, and 10 respectively. We trained the network with data sets of size 7,702 and 16,270 for the excess-background and standard cases respectively. Note that these data sets are slightly larger than the ones reported in Section~\ref{sec:sim_setup} because they include parameters with values outside the prior ranges specified in Section~\ref{sec:sim_setup}. In particular, we include lower and higher values of $\tau$ (between $0.022$ and $0.11$), higher values of $f_X$ (up to 1000) and higher values of $A_{\rm r}$ (up to $\sim 200$\% of the CMB at 1.42 GHz). These extreme models help to train ANN but are not used in the likelihood calculations. To prevent over fitting, we validated the network with 25\% of the total models.

In total, we have used four different emulators that are listed in Table~\ref{tab:ann}. Figure~\ref{fig:ANNerror} shows the excess background emulator accuracy for four $k$-bins. The accuracy of the emulators is quantified through root-mean-square error ($\sigma_{\rm ann}$). Although the plot shows $\sim 20\%$ scatter, we have checked that  $\sigma_{\rm ann}$ is smaller than 10\%, which is our floor on the theoretical uncertainty, $\sigma_{\rm th}(k_c)$. We have also checked the effects of modelling uncertainty on our limits and confirm that the results are mainly dominated by the measurement errors. The effect of the uncertainty in theory and ANN are sub-dominant. In particular, taking the theoretical+ANN error to be 30\% does not change the result. The ANN were used to populate the  parameter space described in Section~\ref{sec:sim_setup}. These emulated models are combined with the original set to comprise the data used for our likelihood analysis.

\vfill
\bsp
\label{lastpage}
\end{document}